\documentstyle[aaspp4,10pt,epsf]{article}
\def\sun{\odot}
\def\lsim{\, \lower2truept\hbox{${< \atop\hbox{\raise4truept\hbox{$\sim$}}}$}\,}
\def\gsim{\, \lower2truept\hbox{${> \atop\hbox{\raise4truept\hbox{$\sim$}}}$}\,}

\def\oneskip{\vskip\baselineskip}
\def\oneskip{\vskip\baselineskip}

\begin{document}

%\today

\title{Physical Conditions and Star Formation Activity in the Intragroup
Medium of Stephan's Quintet\footnote{The National 
Radio Astronomy Observatory is a facility of the
 National Science Foundation operated under cooperative agreement by
 Associated Universities, Inc.}$^,$
\footnote{Based on observations made with ISO, an ESA project with
instruments funded by ESA Member States and with the 
participation of ISAS and NASA.}$^,$
\footnote{Based on observations obtained at
the Hale Telescope, Palomar Observatory as part of a continuing
collaboration between the California Institute of Technology, NASA/JPL,
and Cornell University.}
}

\author{C. K. Xu, N. Lu}
    \affil{Infrared Processing and Analysis Center, Jet Propulsion Laboratory, 
    Caltech 100-22, Pasadena, CA 91125}
\author{J. J. Condon}
\affil{National Radio Astronomy Observatory\altaffilmark{1}, 520
Edgemont Road, Charlottesville, VA 22903\\Electronic mail: 
jcondon@nrao.edu}
\author{M. Dopita}
\affil{Research School of Astronomy \& Astrophysics,
The Australian National University, Cotter Road, Weston Creek ACT 2611,
Australia}
\author{R.J. Tuffs}
\affil{Max-Planck-Institut f\"ur Kernphysik,
    Postfach 103980, D69117 Heidelberg, Germany}
%\received{}
%\accepted{}

\received{2003 February 24}
\accepted{2003 June 10}

\begin{abstract}
New multi-band observations of the famous compact group of galaxies
Stephan's Quintet (SQ) are presented and analyzed.  These include far infrared
(FIR) images at 60$\mu m$ and 100$\mu m$ (ISOPHOT C-100 camera), radio
continuum images at 1.4 GHz (VLA B-array) and 4.86 GHz (VLA C-array), and
long-slit optical spectrographs (Palomar $200"$ telescope).  With
these new data, we aim to learn more about the X-ray/radio ridge in
the middle of the intragroup medium (IGM) and the IGM starburst SQ-A,
both are likely to be caused by the high speed collision ($\sim 900$
km s$^{-1}$) between the intruder galaxy NGC~7318b ($v = 5700$ km s$^{-1}$) and
the IGM ($v = 6600$ km s$^{-1}$). We found that the radio ridge has a
steep nonthermal spectral index ($\alpha = 0.93\pm 0.13$) and an
extremely low FIR-to-radio ratio index ($q < 0.59$).  Its IR emission
can be explained in terms of collisional heating of dust grains by
shocked gas.  The minimum-energy magnetic field strength is $H_{\rm
min} \approx 10\,\mu$G.  The long-slit spectra of sources in the ridge
have typical emission line ratios of shock-excited gas. The very broad
line widths ($\geq 1000$ km s$^{-1}$), and the fact that in some cases more
than two velocity systems were detected along the same line of sight,
provide further evidence for an ongoing collision along the ridge.
The IGM starburst SQ-A has a radio spectral index $\alpha =0.8\pm 0.3$
and a FIR-to-radio ratio index $q=2.0\pm 0.4$, consistent with those of
star forming regions.  The optical spectra of two sources in this
region, M1 ($v=6600$ km s$^{-1}$) and M2 ($v=6000$ km s$^{-1}$), have
typical line ratios of HII regions. Both M1 and M2 have metallicity
slightly higher than the solar value. The star formation rate (SFR)
estimated from the extinction-corrected H$_\alpha$ luminosity of SQ-A
is 1.45 M$_\sun$ yr$^{-1}$, of which 1.25 M$_\sun$ yr$^{-1}$ is due to
the $v=6600$ km s$^{-1}$ component and 0.20 M$_\sun$ yr$^{-1}$ to
the $v=6000$ km s$^{-1}$ component.
\end{abstract}


\section{Introduction}
There is a resurgent enthusiasm recently
on the investigations of the historically famous 
Stephan's Quintet (hereafter SQ), a multi-galaxy system 
discovered by Stephan in late 19th century
(Stephan 1887). As summarized in Sulentic et al.
(2001), many new observational data have
recently become available for this intriguing
source. In particular, a few new frequency windows have
been opened to astronomical observations in this space era. 
The X-ray observations using ROSAT (Sulentic et al. 1995;
Saracco \& Ciliegi 1995; Pietsch et al. 1997; Sulentic et al.
2001) and Chandra (Trinchieri et al. 2003), 
infrared observations using ISO (Xu et al. 1999; Sulentic
et al. 2001), and very high angular resolution 
optical images using HST (Gallagher et al. 2001) have
revealed intriguing new features in this fascinating
galaxy group. Particularly, an unusual IR source 
was discovered by Xu et al. (1999) in the
intragroup medium (IGM) of SQ, located  more than 20 kpc
away from any neighboring galaxy centers. The IR
source, named Source A (hereafter SQ-A) 
by Xu et al. (1999), is apparently associated with 
a starburst triggered by the high speed collision (relative
velocity $\sim$1000 km s$^{-1}$) between the intruder galaxy
NGC~7318b and the IGM. It is the location in the IGM 
that makes the starburst so unusual. This is also the first example
of an ongoing starburst triggered by collisions of 
such high speed, though evidence for a post-starburst was
found previously by Kenney et al (1995) in the high speed
colliding system NGC~4438/4439.

Another outstanding feature in SQ is the large scale shock front
($\sim 40$ kpc) in the IGM between NGC~7319 and NGC~7318b, first
discovered by Allen \& Hartsuiker (1973) as an radio emission ridge in
the 21cm Westerbork radio continuum map, and later confirmed by the
VLA observations of van der Hulst \& Rots (1981) and of
Williams et al. (2003). The high resolution
X-ray maps of SQ (Pietsch et al. 1997; Sulentic et al.
2001; Trinchieri et al. 2003) show an
X-ray emission ridge at almost exactly the same position, making it
very certain that this ridge is the signature of a shock front.  Given
the fact that the relative velocity between the gas-rich intruder
galaxy NGC~7318b and the rest of the group is $\sim 900$ km s$^{-1}$,
and that there is widely spread cold HI gas in the IGM (Shostak et
al. 1984; Williams et al. 2002), it is indeed expected that such a
shock must be happening in the region where the cold gas associated
with the intruder collides with the cold gas in the IGM.

Xu et al. (1999) argued strongly that the two phenomenal events,
namely the ongoing IGM starburst and the large scale shock, are
intimately related with each other: the starburst is very likely
triggered by the same collision that triggered the shock.  In this
paper, we present new observations in the infrared, radio continuum,
and optical spectroscopy for SQ. With these new data and data found in
the literature, we aim to better constrain the physical conditions in
the IGM, particularly in the shock front region and in SQ-A, and
investigate further the physical mechanism linking the two events.
The paper is organized as follows: after this introduction, the new
observations are presented in Section 2 through Section 4.  In Section
5, we study the physical conditions in the IGM. In Section 6 we
investigate the triggering mechanism for the IGM starburst SQ-A. The
Section 7 is devoted to a discussion, and the conclusions are
presented in Section 8. We assume that the distance of
SQ is 80 Mpc, and the distance of the foreground interloper NGC~7320
is 10 Mpc.

\section{Mapping of IR emission with ISO}
\subsection{Observations}
The mapping observations of SQ were carried out 
on May 23, 1996 using ISOCAM and ISOPHOT on board ISO 
(Kessler et al. 1996). 
The total on-target time
for the ISOCAM maps (at 15$\mu m$ and 11.4$\mu m$) 
is 33 ks, and for the ISOPHOT
maps (at 60$\mu m$ and 100$\mu m$) 
is 14 ks. These maps are among the most sensitive
observations ever made by ISO.
The 15$\mu m$, 11.4$\mu m$ and 60$\mu m$ maps
have already been published in Xu et al. (1999) and in
Sulentic et al. (2001). In this paper we present
the 100$\mu m$ map,
which is the last product from our ISO project.
In Table 1, we summarize the ISO observations.

\vskip1truecm

\centerline{{\bf Table 1. Summary of ISO Observations of Stephan's Quintet}}
\oneskip
%\begin{deluxetable}{ccccccccc}
\begin{footnotesize}
\hskip-1.2truecm\begin{tabular}{|c|c|c|c|c|c|c|c|c|c|c|}\hline
camera & array & filter & $\lambda$ & $\delta\lambda$ & detector &
samp. & map & rms & rms & calib.\\
& size & & ($\mu m$) & ($\mu m$) & pixel & step & size & (MJy/sr) & 
 (mJy/beam$^{\dagger}$) &err. \\ \hline
ISOCAM  & 32$\times 32$ & LW8 & 11.4 & 2.3 & $6''\times 6''$ &  $6''\times 6''$ & $5'\times 12'$ & 0.083 & 0.22 & $\sim$20\% \\\cline{2-11}
 -LW & 32$\times 32$ & LW3 & 15.0 & 6.0 & $6''\times 6''$ &  $6''\times 6''$ & $5'\times 12'$ & 0.047 & 0.12 & $\sim$20\% \\  \hline
ISOPHOT & 3$\times 3$ & C60 & 60.8 & 23.9 & $45''\times 45''$ &  $15''\times 15''$ & $6'\times 13'$ & 0.33 & 14 &$\sim$20\% \\ \cline{2-11}
-C100 & 3$\times 3$ & C100 & 103.5 & 43.6 & $45''\times 45''$ &  $15''\times 15''$ & $6'\times 13'$ & 0.40 & 16 & $\sim$20\% \\ \hline
\end{tabular}

$^{\dagger}$ Beam areas are calculated assuming Gaussian beams of
FWHM of 10$''$ for the ISOCAM maps (under-sampled) and 
FWHM of 40$''$ for the ISOPHOT maps (over-sampled).
\end{footnotesize}
%\clearpage
\vskip1truecm

Both the 100$\mu m$ and the 60$\mu m$ maps were taken
using the ISOPHOT C-100 camera in the observation
mode P32 (oversampling mapping mode). The ISOPHOT P32 mode combines the 
standard spacecraft raster pointing mode (stepping in spacecraft 
Y and Z coordinates) with scans in Y made with the 
focal plane chopper (Tuffs \& Gabriel 2003; see also
Tuffs et al. 2002a,b). 
At each spacecraft raster position the 
focal plane chopper is stepped at intervals 
of one third of the detector pixel pitch, resulting
in sky samplings in Y of $\sim\,15\,^{\prime\prime}$ 
for the C100 detector, which is close to the 
limit for Nyquist sampling of 
(17$\,^{\prime\prime}\times\lambda$)/100$\,{\mu}$m. 
The oversampling factor of 3 was chosen
for both maps (sampling step of $\sim 15''$ in
both in-scan and cross-scan directions). Both maps 
have 16 map lines (scans) and 9 pointings per map line,
covering about 13$'$ (in scan) $\times$ 6$'$ (cross scan)
sky area (Table 1). The rational of choosing
relatively long scans (13$'$ compared
to the size of SQ of $\sim 5'$) is two-folds:
(1) to facilitate the correction of the 
transient response behavior of the Ge:Ga photoconductor 
detectors of ISOPHOT; (2) to enclose enough
empty background sky regions surrounding the system
and therefore the maps will be sensitive to any
tenuous diffuse emission associated with the intragroup
medium outside main bodies of member galaxies.

The ISOPHOT data were reduced 
using the latest version of the 
P32 data reduction package (Tuffs \& Gabriel 2003). 
This software takes the advantage of high sampling rate
and high redundancy of P32 maps. 
The effects of detector transient were
properly corrected (failure 
to correct for these effects in data taken in the ``P32'' mode can give 
rise to serious signal losses and distortions in the derived brightness 
profiles through the maps). Detailed descriptions of the software
and of the standard data reduction procedure can be found in Tuffs \& Gabriel
(2003). During the reduction procedure it became clear that,
for the 100$\mu m$ map, three (pixel 1, 4 and 5)
of the nine detector pixels of the C-100 camera
are rather noisy. These pixels were therefore 
excluded from the final map making.

The final calibrations for both the 100$\mu m$ and 60$\mu m$
were determined using a tool provided by
the P32 data reduction package, which compares
the local background with 
that measured by IRAS and COBE, taking also into account the 
seasonal variation of the zodiacal light.
 As pointed out by Tuffs \& Gabriel (2003), this procedure can
     remove any residual systematic uncertainties remaining after the
     transient response correction, and yield close agreements
     with IRAS calibrations.
At 60 and 100$\mu m$, brightnesses on the ISO maps were multiplied
    by factors of 1.04 and 0.81, respectively, to bring the background
    level on the ISOPHOT maps into equivalence with the background estimated
    from COBE/DIRBE.
Aperture photometry using the final 100$\mu m$ and
60$\mu m$ maps finds total flux densities (and
rms errors) of SQ of
$f_{100\mu m}=2.22 (\pm 0.03)$~Jy and $f_{60\mu m}=1.07 (\pm 0.03)$~Jy,
in good agreement with the IRAS values of 
$f_{100\mu m}=2.54\pm 0.36$~Jy and $f_{60\mu m}=0.88 \pm 0.09$~Jy.
From these comparisons, the
    IRAS values are consistent with the ISO values to within 20
    percent. In what follows we will adopt a conservative estimate
    of 20 percent for the systematic calibration uncertainty of
    the ISO data.
%\clearpage

%\vskip4truecm

\begin{figure}
\vbox {
  \begin{minipage}[l]{1.0\textwidth}
   \hbox{
      \begin{minipage}[l]{1.0\textwidth}
       \vspace{-1truecm} 
       {\centering \leavevmode \epsfxsize=\textwidth 
%        \epsfbox{fig1a_red.ps}}
       \vspace{2truecm}}
      \end{minipage} \  \hfill \
   }
  \end{minipage} \  \hfill \
  \begin{minipage}[l]{1.0\textwidth}
   \hbox{
      \begin{minipage}[l]{1.0\textwidth}
    \vspace{1truecm} 
       {\centering \leavevmode \epsfxsize=\textwidth 
%        \epsfbox{fig1b_red.ps}}
       \vspace{2truecm}}
      \end{minipage} \  \hfill \
  }
  \end{minipage} \  \hfill \
\begin{minipage}[l]{1.0\textwidth}
\vspace{0.5truecm} 
\caption{ISOPHOT images of Stephan's Quintet.}
\end{minipage}
}
%\label{fig1}
%\vspace*{-0.2cm}
\end{figure}
%\vskip1truecm

\begin{figure}
\vbox {
  \begin{minipage}[l]{1.0\textwidth}
   \hbox{
      \begin{minipage}[l]{0.6\textwidth}
       \vspace{0truecm} \hspace{0.5truecm}
       {\centering \leavevmode \epsfxsize=\textwidth 
       \vspace{2truecm}}
%        \epsfbox{fig2a.ps}}
      \end{minipage} \  \hfill \
      \begin{minipage}[r]{0.6\textwidth}
       \vspace{0truecm} \hspace{-1.5truecm}
       {\centering \leavevmode \epsfxsize=\textwidth 
       \vspace{2truecm}}
%        \epsfbox{fig2b.ps}}
      \end{minipage} \  \hfill \
   }
  \end{minipage} \  \hfill \
  \begin{minipage}[l]{1.0\textwidth}
   \hbox{
      \begin{minipage}[l]{0.6\textwidth}
    \vspace{1truecm}  \hspace{0.5truecm}
       {\centering \leavevmode \epsfxsize=\textwidth 
       \vspace{2truecm}}
%        \epsfbox{fig2c.ps}}
      \end{minipage} \  \hfill \
      \begin{minipage}[r]{0.6\textwidth}
       \vspace{1truecm} \hspace{-1.5truecm}
       {\centering \leavevmode \epsfxsize=\textwidth 
       \vspace{2truecm}}
%        \epsfbox{fig2d.ps}}
      \end{minipage} \  \hfill \
  }
  \end{minipage} \  \hfill \
\begin{minipage}[l]{1.0\textwidth}
\vspace{0.5truecm} 
\caption{Contour maps of the IR emission (background
subtracted) at 100$\mu m$, 60$\mu m$, 15$\mu m$, and 11.4$\mu m$
overlaid on the 15$\mu m$ image.  The contour levels for the 100$\mu
m$ map are [2,2.8,4,5.6,8,11.2,16]$\times$0.4 MJy/sr; for the 60$\mu m$ map
they are [2,2.8,4,5.6,8,11.2,16,22.4]$\times$0.33 MJy/sr; for the 15$\mu m$
map they are [2,2.8,4,5.6,8,11.2,16,22.4,32]$\times$0.047 MJy/sr; for the
11.4$\mu m$ map they are [2,2.8,4,5.6,8,11.2,16]$\times$0.083 MJy/sr.
Sources detected in the 15$\mu m$ map are labeled according to 
Xu et al. (1999).}
\end{minipage}
}
%\label{fig2}
%\vspace*{-0.2cm}
\end{figure}
%\vskip1truecm

\subsection{Results}
In Fig.~1a and Fig.~1b we present the images of the
100$\mu m$ and 60$\mu m$ maps, respectively. In both maps the background
is quite smooth and the noise behaves normally.
No obvious artifacts can be seen in the images except for near
the edges. 
The emission in both maps is dominated by two sources,
one associated with the Seyfert galaxy
NGC~7319 (on the right), the other with the foreground galaxy
NGC~7320. These two galaxies were marginally resolved and
detected in the 60$\mu m$ IRAS HIRES map (Allam et al. 1996).
In other three IRAS bands (12, 25, and 100$\mu m$), 
only NGC~7319 was detected in the IRAS HIRES study of 
Allam et al. (1996). 

In both ISOPHOT 100$\mu m$ and 60$\mu m$ maps,
the source associated with NGC~7319 
is slightly elongated along the scan direction
(P.A.= 167$^\circ$), with the size (FWHM) along the major axis 
being 48$''$ and 54$''$, respectively, and
the minor/major axis ratio being 0.78 and 0.61, respectively.
This might indicate that, in addition to the IR emission of the
Sy2 nucleus which dominates the MIR emission of 
NGC~7319 (Xu et al. 1999; Sulentic et al. 2001),
the dust in the host galaxy 
may contribute significantly to the FIR emission.
The H$_\alpha$ arm on the north of the nucleus
(Xu e al. 1999; see also Fig.~6 below) where massive
molecular gas was also detected (Yun et al. 1997; Gao \& Xu 2000),
may indeed be active star formation regions in 
the disk of NGC~7319 and therefore be bright in 
the FIR. On the other hand, 
it cannot be ruled out that the elongation is caused
by some residual transient effects in the ISO maps,
given the coincidence with the scan direction. Future
observations with higher angular resolutions (e.g. using
SIRTF-MIPS) will help to
distinguish these two possibilities.

In Fig.~2a, 2b, 2c and 2d, 
contours of the 100$\mu m$, 60$\mu m$, 15$\mu m$,
and 11.4$\mu m$ maps are overlaid on the 15$\mu m$
image, respectively. 
Sources A, B, C, D, together with
member galaxies of SQ, are marked on Fig.~2c (the 15$\mu m$
image, see also Fig.~ of Xu et al. 1999).
Comparisons with ISOCAM maps
show that the IGM starburst SQ-A, as well as
the main body of the galaxy pair NGC~7318a/b are 
clearly detected in the longer wavelength ISOPHOT maps.
Another weaker IGM starburst discovered in
the ISOCAM 15$\mu m$ map (Xu et al. 1999), 
Source B (hereafter SQ-B), also 
appears to be detected in both the 100$\mu m$ and
the 60$\mu m$ maps. However, its ISOPHOT
flux densities 
are highly uncertain (at about a factor of 2 level) because
SQ-B is located on a background plateau 
due to a more extended, at same time rather structured,
low-emission region and therefore it is difficult  
to subtract the background accurately.

It is always difficult to measure flux densities for 
individual sources in a crowded field such as SQ.
This problem becomes even more severe in the ISOPHOT maps
where the angular resolutions are poor ($\sim 40''$). 
Indeed, the ISOPHOT flux densities of the weaker sources such as SQ-A
and NGC~7318a/b had to be measured on the residual 
maps after subtracting 
the two bright sources NGC~7319 and NGC~7320.
In this procedure, the two bright sources 
are approximated by two 2-dimensional Gaussians
in each of the ISOPHOT maps, and
are then subtracted from the map.
In Fig.~3a and Fig.~3b, we present the contours 
of the residual 100$\mu m$ map and 60$\mu m$
map, respectively. Given the surface brightness distributions
of NGC~7319 and of NGC~7320 in the higher resolution ISOCAM maps,
the 2-dimensional Gaussian model should be a good
approximation. Nevertheless, the ISOPHOT flux densities of 
SQ-A and NGC~7318a/b have large uncertainties due to
the sensitive dependence on the subtraction of brighter sources.

In Table 2, we list the IR flux densities of the individual
sources measured from the ISO maps. 
Except for the flux densities 
presented with brackets denoting the high uncertainties
(see discussion above), the errors of
the measurements are on the order of 20\%, mostly due
to the adopted (rather conservative) calibration 
uncertainty (see Section 2.1). The errors of those
flux densities with brackets can easily be 50\% or more.
The coordinates of the
galaxies are taken from NED, while the positions
of the ISOCAM sources (Source A through D) are
measured from the ISOCAM 15$\mu m$ image, with uncertainties
on the order of 2$''$.

In the 100$\mu m$ map,
there is clear evidence for extended emission 
(at $\sim 1$ MJy/sr level) in the periphery
of SQ, with a broad wing passing through SQ-B.
Since much of this region is not in the same scans 
that pass though the bright sources, the diffuse
emission is unlikely to be due to any residual 
transient effects. An estimate of the
 integrated flux densities of the diffuse emission is given by
 subtracting the sums of the flux densities of the discrete sources
 listed in Table 2 (respectively 0.68 and 1.32 Jy at 60 and 100$\mu m$),
 from the total flux densities of SQ (equal to 1.07 and 2.22 Jy
 at the same two wavelengths). This yields values of 0.39 and 0.90 Jy
 for the diffuse component at 60 and 100$\mu m$, respectively,
 which is about 40 percent of the total.
This is one of a few examples where diffuse dust emission
 is detected outside galaxies (see Tuffs \& Popescu 2002 for a
 review on this topic). The diffuse emission around SQ-B is likely
 to be powered by young stars formed in the same tidal tail
 where SQ-B is located. Braine et al. (2001) and Lisenfeld 
et al. (2002) detected
 massive molecular gas in this region. In other places,
 it might be powered either by stars stripped from galaxies
 during previous close encounters between SQ members
 (Moles et al. 1997; Sulentic et al. 2001), or by the diffuse
 X-ray emitting gas (Trinchieri et al. 2003). This will be a very
 interesting problem for future, more sensitive IR observations.

The early type galaxy NGC~7317, detected by
ISOCAM at both 15$\mu m$ and 11.4$\mu m$, was
not detected by ISOPHOT, nor were ISOCAM sources 
Source C (a star in the north of SQ) and Source D 
(a background galaxy in the south of SQ).

%
%\clearpage
\vskip3truecm
\centerline{{\bf Table 2. ISO Flux Densities of Sources in
SQ Field}}
\oneskip\nobreak
%\begin{deluxetable}{ccccccccc}
\hskip-1truecm\begin{tabular}{|l|l|l|l|l|l|l|l|l|l|l|l|}\hline
Name  & R.A. & Dec. &  
$f_{11.4\mu m}$ & $f_{15\mu m}$ & $f_{60\mu m}$ & $f_{100\mu m}$ & note \\
 & (J2000) & (J2000) & (mJy) & (mJy) & (mJy) & (mJy) & \\\hline
NGC~7317 & 22~35~52.0 &  +33~56~41 & 1.6 & 2.2 & & & E galaxy\\\hline
Shock \& NGC~7318a/b$^\dagger$ &  22~35~57.6 & +33~57~57 & 23.9 & 19.2 & (76)
& (230) & shock \& g. pair \\\hline
NGC~7319 & 22~36~03.5 & +33~58~33 & 54.5 & 76.3 & 275 & 350 & Sy2 galaxy \\\hline
NGC~7320 &  22~36~03.5 &  +33~56~54 & 46.7 & 27.3 & 260 & 609 & f.g. galaxy \\\hline
SQ-A$^\ddagger$ & 22~35~58.7 & +33~58~50 & 10.9 & 10.9 & 
(60) & (106) & IGM starburst \\\hline
SQ-B &  22~36~10.2 &+33~57~21 & 0.9 & 1.7 & (13) & (21) & IGM starburst\\\hline
SQ-C &  22~36~02.7 &+33~59~56 &3.3 & 0.6 & & & star \\\hline
SQ-D &  22~36~00.1 &+33~55~52 & (0.2) & 0.7 & & &b.g. galaxy\\\hline
\end{tabular}

$^\dagger$ These 3 regions are grouped together because 
they can hardly be separated from each other in the ISOPHOT
maps (Fig.~3).

$^\ddagger$ The ISOPHOT flux densities
of SQ-A were measured in the region
outlined by the northwest peak in the residual 60$\mu m$
map (Fig.~3b).

\vskip1truecm

\begin{figure}[h!]
\vbox{
  \begin{minipage}[l]{1.0\textwidth}
   \hbox{
      \begin{minipage}[l]{0.6\textwidth}
       \vspace{0truecm} \hspace{0.5truecm}
       {\centering \leavevmode \epsfxsize=\textwidth 
       \vspace{2truecm}}
%        \epsfbox{fig3a.ps}}
      \end{minipage} \  \hfill \
      \begin{minipage}[r]{0.6\textwidth}
       \vspace{0truecm} \hspace{-1.5truecm}
       {\centering \leavevmode \epsfxsize=\textwidth 
       \vspace{2truecm}}
%        \epsfbox{fig3b.ps}}
      \end{minipage} \  \hfill \
    }
  \end{minipage} \  \hfill \
\begin{minipage}[l]{1.0\textwidth}
   \vspace{0.5truecm} 
\caption{Contour maps of the 100$\mu m$ and
60$\mu m$ emission after the subtraction
of NGC~7319 and NGC~7320, 
overlaid on the 15$\mu m$ image.
The contour levels for the 100$\mu m$ map
are [2,2.8,4,5.6]$\times$0.4 MJy/sr;
for the 60$\mu m$ map they
are [2,2.8,4,5.6]$\times$0.33 MJy/sr.
}
\end{minipage}
}
%\label{fig3}
%\vspace*{-0.2cm}
\end{figure}
%
%\vskip1truecm

\section{Radio Continuum Observations with the VLA}
\subsection{Observations and Data Reduction}
There have been several interferometric radio continuum observations
of SQ in the literature (Allen \& Hartsuiker 1973; 
van der Hulst \& Rots 1981; Williams et al. 2002;
Xanthopoulos et al. 2002). With new high sensitivity,
high angular resolution VLA imaging observations in two bands
(1.40 and 4.86 GHz), we aim to constrain the spectral indices
of individual sources. We also try to detect the polarization,
particularly in the shock front region, 
in an attempt to constrain the magnetic field.

In order to image the radio continuum brightness, polarization, and
spectral index distributions of SQ, we observed SQ with the VLA at
1.4GHz in B configuration for 5 hrs on 1999 November 15 and at 4.86GHz
in C configuration for 8 hrs on 2000 April 27.  During both
observations, the source 3C~48 ($S_{1.40\,{\rm GHz}} = 16.19$~Jy,
$S_{4.86\,{\rm GHz}} = 5.52$~Jy, Baars et al.~1977) was used to
calibrate the fringe amplitudes to the updated VLA flux-density scale.
The polarization position angles were determined from 3C~138, and the
nearby source J2236+284 was used as the phase calibrator.

Preliminary total-intensity (Stokes I) images were made and cleaned by
the AIPS task IMAGR.  To maximize surface-brightness sensitivity and
in the same time minimize the synthesized sidelobes, 
we used tapered uniform weighting with
Gaussian amplitude tapering to 30\% at 25,000 wavelengths and
truncated at 45,000 wavelengths. 
This strong taper yields a large synthesized beam area having the high
surface-brightness sensitivity of natural weighting.  The nearly
uniform weighting below 25,000 wavelengths reduces the dirty-beam
sidelobes that natural weighting would produce because the central
part of the VLA synthetic aperture would be "over-illuminated."
 Then the clean components from these
images were subtracted from the $(u,v)$ data.  All residual
visibilities having amplitudes much larger than the rms noise were
flagged, and the clean components were added back to the edited data
set.  This procedure removes low-level interference and other problems
that cause individual visibilities to disagree significantly with the
data set as a whole.  New images of 1024 pixels $\times 1''$ pixel$^{-1}$
on a side were made, cleaned, and restored with $6\,\farcs0$ FWHM
circular Gaussian beams, and the final images were corrected for
primary-beam attenuation.  The rms noise levels are $\sigma \approx
25\,\mu$Jy~beam$^{-1}$ at 1.40~GHz and $\sigma \approx
17\,\mu$Jy~beam$^{-1}$ at 4.86~GHz.  Matching Stokes Q and U images
were also made, and they have somewhat lower noise levels.  No
linearly polarized emission brighter than 50~$\mu$Jy~beam$^{-1}$ was
found in any component of SQ at either 1.40 or 4.86~GHz.

\subsection{Radio Continuum Images} 

In Fig.~4 we present the total-intensity contour maps of
the 1.40 GHz and 4.86 GHz images, both overlaid on an R-band
CCD image (Xu et al. 1999).
%\vskip1truecm
\begin{figure}
\vbox{
  \begin{minipage}[l]{1.0\textwidth}
   \hbox{
      \begin{minipage}[l]{0.5\textwidth}
       \vspace{0truecm} \hspace{-0.4truecm}
       {\centering \leavevmode \epsfxsize=\textwidth 
       \vspace{2truecm}}
%        \epsfbox{fig4a.ps}}
      \end{minipage} \  \hfill \
      \begin{minipage}[r]{0.5\textwidth}
       \vspace{0truecm} \hspace{-0.4truecm}
       {\centering \leavevmode \epsfxsize=\textwidth 
       \vspace{2truecm}}
%        \epsfbox{fig4b.ps}}
      \end{minipage} \  \hfill \
    }
  \end{minipage} \  \hfill \
\begin{minipage}[l]{1.0\textwidth}
%\vspace{-1.5truecm} 
\caption{{\bf Left}: Contours of the 
radio continuum at 1.40 GHz (VLA B-array) overlaid on an R-band
CCD image. 
The lowest contour is 90~$\mu$Jy~beam$^{-1}$ and
the spacing $= 2$ in ratio. {\bf Right}: Contours of the 
radio continuum at 4.86 GHz (VLA C-array) overlaid on the same R-band
image. The lowest contour is 50~$\mu$Jy~beam$^{-1}$ and
the spacing $= 2$ in ratio. For both radio maps the
FWHM of the synthesized beam is $6\,\farcs0$.
}
\end{minipage}
}
%\label{fig4}
%\vspace*{-0.2cm}
\end{figure}

The following radio sources are visible in these maps and their fitted
parameters are listed in Table 3:
\begin{description}
\item{(1)} There is a strong steep-spectrum source whose centroid at
J2000 $\alpha = 22^{\rm h}~36^{\rm m}~03\,\fs55$, $\delta =
+33^\circ~58'~32\,\fs6$ overlaps the optical nucleus of the Seyfert 2
galaxy NGC~7319 at $\alpha = 22^{\rm h}~36^{\rm m}~03\,\fs56$, $\delta
= +33^\circ~58'~33\,\farcs2$.  We fit this source with elliptical
Gaussians and deconvolved the $6\thinspace\farcs0$ restoring beams to
obtain Gaussian approximations to the source FWHM major axis, minor
axis, and position angle at both frequencies: $4\thinspace\farcs5
\times 1\thinspace\farcs6$ in PA $= 29^\circ$ E of N in the 1.40
GHz map, $4\thinspace\farcs5 \times 1\thinspace\farcs5$ in PA $=
27^\circ$ in the 4.86 GHz map.  These sizes and orientations are
reliable even though the deconvolved source size is smaller than the
beam because the signal-to-noise ratio of this source is so high in
our images.  They are consistent with the linear triple structure
found in higher-resolution VLA images at 3.6, 6, and 20~cm (Aoki et
al.~1999).  The precise agreement of our 1.40~GHz integrated flux
density with theirs (28.5~mJy in both cases) indicates that there can
be very little extended emission surrounding the triple.  The
orientation of the radio triple is nearly parallel to the
H$_\alpha$-[N~II] emission-line region visible in Fig.~2 of Xu et
al.~(1999).  This phenomenon is common in Seyfert galaxies and
suggests that the radio jets are brightening the line-emitting gas by
compression.
\item{(2)} There is a weak radio source in NGC~7318a, the
early type galaxy in the binary system NGC~7318a/b.  This radio source
is unresolved (FWHM $< 4''$) in our images and centered on $\alpha =
22^{\rm h}~35^{\rm m}~56\thinspace\fs72$, $\delta =
+33^\circ~57'~56\thinspace\farcs0$, precisely coincident with the
optical nucleus at $\alpha = 22^{\rm h}~35^{\rm
m}~56\thinspace\fs74$, $\delta  = +33^\circ~57'~56\thinspace\farcs3$
(Klemola et al. 1987). The source has been detected previously
by van der Hulst \& Rots (1981) and by Williams et al. (2002)
at 1.4 Ghz. Our measurement of the $S_{1.40GHz}$ (0.95$\pm 0.05$ mJy)
is slightly ($\sim 2sigma$) lower than that of Williams et al.
(1.4$\pm 0.2$).  
\item{(3)} There is a radio ridge at $\alpha \approx 22^{\rm
h}~36^{\rm m}~00^{\rm s}$, which extends north-south from $\delta
\approx +33^\circ~57'~10''$ to $\delta \approx +33^\circ~58'~40''$,
then bends west and terminates near $\alpha = 22^{\rm h}~35^{\rm
m}~56^{\rm s}$, $\delta = +33^\circ~59'~15'$ 
(Allen \& Hartsuiker, 1973; van der Hulst \& Rots, 1981;
Williams et al. 2002). The radio ridge is
coextensive with the X-ray ridge (Pietsch et al. 1997), the ridge of
the high-redshift (6600 km~s$^{-1}$) component of the H$_\alpha$-[N~II]
emission (Xu et al. 1999), and the FIR emission shown in Fig.~3.
With a very faint optical continuum counterpart, the ridge
appears to delineate a shock front containing relativistic electrons,
magnetic fields, hot thermal electrons, ionized gas, and cool dust.
The 1.4 GHz flux density, $S_{1.40GHz} = 34.7\pm 3.5$ mJy, 
is about 30\% lower than
the result of Williams et al. (2002) obtained using their 
lower resolution image
which is $S_{1.40GHz} = 48\pm 7$ mJy, indicating that 
our high-resolution B-array image may be missing about 13 mJy of flux from the
ridge.
\item{(4)} The relatively weak and diffuse radio peak
near $\alpha = 22^{\rm h}~35'~ 58\,\fs8$, $\delta = +33^\circ~58'~50''$
overlaps the IGM starburst SQ-A just north of the bend in the
ridge.  Its 1.40 and 4.86 GHz flux densities were estimated
by fitting Gaussians to SQ-A above base planes fitted to
the surrounding ridge emission.
\item{(5)} The binary radio source (called SQ-R in Table 3), 
with the stronger component at $\alpha = 22^{\rm h}~36^{\rm
m}~00\,\fs01$, $\delta = +33^\circ~59'~12\,\farcs3$ and the weaker
component at $\alpha = 22^{\rm h}~35^{\rm m}~59\,\fs9$, $\delta =
+33^\circ~59'~02''$, is 
almost certainly cosmologically distant background source seen in
projection behind SQ (van der Hulst \& Rots, 1981;
Williams et al. 2002). It has no optical counterpart brighter
than 29 mag arcsec$^{-2}$ (Williams et al. 2002), nor any
IR counterpart in the ISO images. The 1.40 GHz flux density
is in excellent agreement with Williams et al. (2002).
\item{(6)} The weak IGM starburst SQ-B
(near the eastern boundary of the images)
is marginally detected in both 1.40 GHz and 4.86 GHz maps.
\end{description}
The flux densities of these sources are listed in Table 3.
Their total 1.40 GHz flux is 76.5 mJy, somewhat smaller than
the $S_{1.4}= 93 \pm 4$ mJy from the D-configuration NVSS image, so our
B-configuration image may be missing up to 16.5 mJy of diffuse flux.
A comparison with the total flux density ($S_{1.4}= 96 \pm 15$ mJy)
from the C and D configurations image of Williams et al. (2002)
leads to a similar conclusion.

%\vskip3truecm
%
\clearpage

\centerline{{\bf Table 3. Radio Continuum Properties of Sources in
SQ Field}}
\oneskip
%\begin{deluxetable}{ccccccccc}
\begin{tabular}{|l|l|l|l|l|c|c|}\hline
Name  & R.A. & Dec. &$S_{\rm 1.40~GHz}$ & $S_{\rm 4.86~GHz}$ & spect. index 
& FIR/radio index \\
 & (J2000) & (J2000) &  mJy & mJy & ($\alpha$) & (q)   \\\hline
Ridge& 22~35~59.8 & +33~58~00 & 34.7$\pm 3.5$ & 10.9$\pm 1.1$ & 0.93$\pm 0.13$  & $<0.59^\dagger$ \\  \hline
NGC~7318a & 22~35~56.72 & +33~57~56.0 &  0.95$\pm 0.05$ & 0.44$\pm
0.03$ & 0.62$\pm 0.07$ & \\ \hline
NGC~7319 & 22~36~03.55 & +33~58~32.6 & 28.5$\pm 0.5$ & 9.7$\pm 0.3$ & 0.87$\pm 0.03$ & 1.08$\pm 0.10$ \\ \hline
SQ-A & 22~35~58.8 & +33~58~50 & 0.8$\pm 0.2$ & 0.3$\pm 0.1$ & 0.8$\pm
0.3$ & 
2.0$\pm 0.4$ \\ \hline
SQ-B & 22~36~10.2 &+33~57~21 & 0.6$\pm 0.2$ & 0.2$\pm0.1$ &
0.7$\pm 0.4$ & 1.5$\pm 0.7$ \\ \hline
SQ-R & 22~36~00.01 & +33~59~12.3 & 10.3$\pm 1.0$ & 3.7$\pm 0.4$ & 0.85$\pm 0.12$  & \nodata \\  \hline
\end{tabular}

$^\dagger$ The $q$ (upperlimit) is calculated using the ISOPHOT
flux densities of a region including also NGC~7318a/b (Table 3).  

\vskip1truecm

\begin{figure}
\vspace{2truecm}
%\plotone{fig5.ps}
\caption{Contours of radio spectral index ($\alpha$) overlaid
on ISOCAM 15$\mu m$ image. The spatial resolution is $10''$ (FWHM).
Only regions having $s/\sigma \geq 2$ in both radio maps
are included in the calculation of $\alpha$.
The contour levels are $\alpha = 0.5$, 0.7, 0.9 and 1.1.}
\end{figure}
%\vskip1truecm

In Fig.~5 we present the contour map of the
spectral index $\alpha = \log(S_{1.4}/S_{4.86})/\log(4.86/1.4)$
overlaid on the ISOCAM 15$\mu m$ map. It should be noted that
there is little effect of the missing flux on the spectral 
index analysis because (1) the $(u,v)$-plane coverages
and synthetic beams of the two images
are very similar and therefore about same fraction of flux is
missed in the two bands, and (2) the flux missed is mostly in
the diffuse emission that contributes little to the flux ratios
of the relatively high surface brightness regions where the
spectral index analysis is confined.
Both the 1.40 GHz and the 4.86 GHz maps were smoothed to a common
Gaussian beam of FWHM=$10''$, the beam size of
the ISOCAM map, before the calculation of the
spectral index. Only pixels with s/$\sigma > 2$ in both maps 
were included in the calculation. Both the background binary
source and the strong source associated
with the Seyfert 2 galaxy NGC~7319 have fairly 
normal spectral indices $\alpha \sim 0.8$, and the radio source
associated with NGC~7318a has a slightly flatter spectral index
 $\alpha \sim 0.6$.
A large part of the radio ridge associated with the shock front has
$0.7 < \alpha < 1.1$, typical for the synchrotron emission from
relativistic electrons, while a small part of the ridge
immediately south of SQ-A shows steep ($\alpha \sim 1.2$) spectrum. It is
interesting to note that, albeit not being plotted by the spectral index
contours in Fig.~5 because of low s/$\sigma$, this 
steep spectrum appears to extend
to the northwest of SQ-A where diffuse X-ray (Trinchieri et al. 2003)
and H$_\alpha$ (Fig.~6) emission was detected. 
Since a steep-spectrum radio emission is usually associated with
aging CR electrons, its distribution may provide constraints 
to the dynamic history of the shock.

\subsection{Comparison with IR Maps}
For normal galaxies there is
a strong correlation between the IR 
and the radio continuum emission, not only for the total
fluxes in these two bands (Helou et al. 1985; De Jong et al.
1985; Popescu et al. 2002) 
but also for the local emission within galaxies,
for example in the LMC (Xu et al. 1992) and in
M31 (Hoernes et al. 1998). The very tight
correlation is generally explained by the
common dependence of the two emissions on
the star-formation activity (see Condon 1992, and
V\"olk \& Xu 1994 for reviews).

However, in the case
of SQ, the IR maps (Fig.~2's) and the radio continuum
maps (Fig.~4) look very different. Also the 
ratio between the total FIR flux and the total 
radio continuum flux of SQ is significantly different
from the typical value for normal galaxies:
The ratio is 
often expressed by the $q$ parameter (Helou
et al. 1985) which will be called the FIR-to-radio ratio index
in this paper:
\begin{equation}
q\equiv\log\left({FIR\over 3.75\times 10^{12}{\rm ~W~m}^{-2}}\right)
   -\log\left({S_{\rm 1.40~GHz}\over {\rm W~m}^{-2}{\rm ~Hz}^{-1}}\right)
\end{equation}
where
\begin{equation}
FIR=1.26\times 10^{-14}\left({2.58f_{60\mu{\rm m}} + f_{100\mu{\rm  m}}
\over {\rm Jy}}\right).
\end{equation}
For SQ ($S_{\rm 1.40~GHz}=93 \pm 10$ mJy, 
$f_{60\mu\,{\rm m}}=1.07\pm 0.21$ Jy and 
$f_{100\mu\,{\rm m}}=2.22 \pm 0.51$ Jy), $q=1.23\pm 0.27$, significantly
lower than the mean for normal star-forming galaxies,
which is $\langle q \rangle = 2.3\pm 0.2$ (Helou et al. 1985).
The $q$ values for individual sources in SQ
are listed in Table 3. 

Noticeably, the upperlimit of the $q$ ($< 0.59$) of the radio ridge
indicates that it is more than a factor of 50 
too radio-loud compared to normal star-forming galaxies. 
This is because the ridge
is not in a galaxy dominated by ongoing star formation.
Rather, most of the radio emission 
is from relativistic electrons accelerated
by the large-scale shock, which
is in turn created by the collision
between the intruder and the IGM (van der Hulst
\& Rots 1981; Pietsch et al. 1997). 
When developing a theory for the break-down
of the FIR/radio correlation in clustered galaxies,
V\"olk \& Xu (1994) considered a scenario
bearing many similarities with what is actually
happening in SQ (though SQ is 
a group, not a cluster). They
envisaged a high speed collision between a galaxy and 
a clump of cold intracluster medium, the latter
being stripped gas (e.g. by ram pressure) 
from a late-type galaxy. The collision produces 
a large amount of cosmic-ray electrons which in turn
radiate strongly in the radio continuum, while
little FIR emission is produced.

Another reason for the low $q$ value 
of SQ is the radio emission from
the AGN in the Seyfert 2 galaxy NGC~7319, which accounts
for about one-third of the emission in SQ in
both the IR and radio wavebands.
NGC~7319 has a $q=1.10$ (Table 3), low compared
to the mean of normal star-forming galaxies but in the
range of those of AGN's (Condon \& Broderick 1988).
The physical mechanisms responsible
for the relation between the radio and the FIR emissions
in AGNs are completely different 
from those in normal star-forming galaxies (Condon \& Broderick 1988).

Given the relatively large uncertainty, the $q$ value
of SQ-A ($q=2.0\pm 0.4$) is consistent with 
normal star formation. For SQ-B, another starburst in the IGM of SQ, 
$q=1.5\pm 0.7$. This is substantially lower than
the mean, though only at $\sim 1 \sigma$ significance level because
of the large error bar. SQ-B has large uncertainties
in both the FIR and the radio continuum flux densities,
mainly because it is located on a background plateau 
due to a more extended, at same time rather structured,
low-emission region. This makes not only the
the background subtraction difficult but also
the boundaries of the emission
associated with this source highly uncertain.
 
\section{Long-Slit Optical Spectroscopy}
\subsection{Observations}
Long-slit optical spectroscopic observations were made on September 21, 1999 (UT)
using the Double Spectrograph mounted on the Cassegrain focus of Palomar 200$''$ 
telescope.  For the red channel (5600{\AA} -- 8000{\AA}), 316 lines/mm grating 
mode (102{\AA}/mm) was chosen, and for the blue channel (3700{\AA} -- 5400{\AA}), 
600 lines/mm grating mode (71{\AA}/mm) was used.  Two $1024\times 1024$ CCD 
detectors, one for each channel, have pixel size of 24$\mu m$.  The slit is 
128$''$ long and 2$''$ wide.  The spectral dispersions for the red and blue spectra 
are 2.4{\AA}/pix and 1.7{\AA}/pix, respectively. Measured from the bright
sky lines, the FWHM spectral resolutions are
7.5{\AA} for the red channel and 4.6{\AA} for the blue.
 These correspond to
velocity resolutions of 343 km s$^{-1}$ at H$_\alpha$ and 283 km s$^{-1}$ at H$_\beta$,
respectively.

%\vskip1truecm
%
\begin{figure}
\vspace{2truecm}
%\plotone{fig6.ps}
\caption{Positions of Slit M and Slit N. Each is  
128$''$ long and 2$''$ wide. The labels ``1'', ``2'', ``3'' and ``4''
along Slit M and ``a'', ``b'', ``c'', ``d'' and ``e'' along
Slit N mark the approximate positions of the regions where
emission lines are detected (Table 4). The background, taken
from Fig.~2 of Xu et al. (1999), is
the K$'$ (2.1$\mu m$) image overlaid by contours of the 
H$_\alpha$--[NII] emission. 
}
\end{figure}

%\vskip1truecm
Observations were carried out along two slits: Slit M and Slit N
(Fig.~6). Slit M is centered at (J2000) 22h35m58.7s 
+33$^\circ$58$'$50$''$, with a position angle of 128 deg.
Slit N is centered at (J2000) 22h35m59.8s 
+33$^\circ$58$'$00$''$, with a position angle of 176 deg.
Three exposures, each lasting for 20 minutes, were carried out for each slit
orientation.  The spectra derived from these individual exposures were coadded 
for each slit orientation during the data reduction which was done using IRAF
developed by National Optical Astronomy Observatories.

The standard neon and hollow cathode lamp spectra were taken before and after 
each source observation for the purpose of wavelength calibration.  For each
red or blue spectrum, about 20 to 25 bright emission lines with accurately 
known wavelengths were used to determine a 
wavelength solution with an accuracy
of better than 0.2{\AA}.  
The flux calibration was done via observations of the standard stars BD$+$28$^o$4211 
and G191B2B from (Oke, 1990) in a $4''$-wide slit.  The night 
was not strictly photometric, with some thin cirrus clouds and a variable seeing
between $1.2''$ to $1.5''$.  While the relative flux calibration across a spectrum
should be quite good, the absolute flux calibration is less certain, probably
not much better than 30\%.
%\vskip1truecm
%
\begin{figure}
\vbox {
%  \vspace{-2.5truecm} 
  \begin{minipage}[c]{1.2\textwidth}
   \hbox{
%      \begin{minipage}[c]{0.7\textwidth}
       {\centering \leavevmode \epsfxsize=\textwidth 
%        \epsfbox{spectrograph.ps}}
       \vspace{2truecm}}
%        \epsfbox{fig7.ps}}
%      \end{minipage} \  \hfill \
  }
  \end{minipage} \  \hfill \
\begin{minipage}[l]{1.0\textwidth}
%\vspace{-3.5truecm} 
\caption{Spectrographs of the IGM star formation regions 
and of the shock front in Stephan's Quintet (see Fig.~6 for
the locations of Slit M and Slit N). 
The approximate positions are marked for redshifted 
emission lines of
[OII]~3727{\AA}, $H_\beta$, [OIII]~4959/5007{\AA},
[OI]~6300{\AA}, $H_\alpha$ (including [NII]~6548/6583{\AA}), 
and [SII]~6717/6731{\AA}. Emission regions along Slit M and
Slit N are labeled (near $H_\alpha$/[NII] lines) 
in the same way as in Fig.~6.}
\end{minipage}
}
\end{figure}
%\vskip1truecm
\setcounter{figure}{7}
\begin{figure}
%\vskip-1truecm
\vspace{4truecm}
%\plotone{fig8.ps}
%\vskip-0.8truecm
\caption{Blue (3700{\AA} -- 5400{\AA}) and red (5600{\AA} -- 8000{\AA})
spectra of 4 sources along Slit M.} 
\end{figure}
%\vskip1truecm
\begin{figure}
%\vskip-1truecm
\vspace{4truecm}
%\plotone{fig9.ps}
%\vskip-0.8truecm
\caption{Blue (3700{\AA} -- 5400{\AA}) and red (5600{\AA} -- 8000{\AA})
spectra of 5 sources along Slit N.} 
\end{figure}

\vskip1truecm
\noindent{\bf Table 4. Spectroscopic Sources along the Two Slits}
\nopagebreak

\hskip-0.5truecm\begin{tabular}{lccc}\hline
Source ID & R.A. (J2000) & Dec. (J2000) & aperture \\ 
\hline
&&& \\
M1 &  22$^{\rm h}$~35$^{\rm m}$~59.18$^{\rm s}$  &    33$^\circ$~58$'$~45.3$''$  &   9$'' \times 2''$ \\
M2 &  22$^{\rm h}$~35$^{\rm m}$~58.11$^{\rm s}$  &    33$^\circ$~58$'$~55.6$''$  &  15$'' \times 2''$ \\
M3 &  22$^{\rm h}$~36$^{\rm m}$~01.90$^{\rm s}$  &    33$^\circ$~58$'$~18.8$''$  &  18$'' \times 2''$ \\
M4 &  22$^{\rm h}$~35$^{\rm m}$~56.54$^{\rm s}$  &    33$^\circ$~59$'$~10.9$''$  &   8$'' \times 2''$ \\
&&& \\
Na &  22$^{\rm h}$~35$^{\rm m}$~59.77$^{\rm s}$  &    33$^\circ$~58$'$~01.8$''$  &  14$'' \times 2''$ \\
Nb &  22$^{\rm h}$~35$^{\rm m}$~59.91$^{\rm s}$  &    33$^\circ$~57$'$~36.8$''$  &   8$'' \times 2''$ \\
Nc &  22$^{\rm h}$~35$^{\rm m}$~59.62$^{\rm s}$  &    33$^\circ$~58$'$~26.2$''$  &   8$'' \times 2''$ \\
Nd &  22$^{\rm h}$~35$^{\rm m}$~59.50$^{\rm s}$  &    33$^\circ$~58$'$~47.9$''$  &   8$'' \times 2''$ \\
Ne &  22$^{\rm h}$~36$^{\rm m}$~00.02$^{\rm s}$  &    33$^\circ$~57$'$~17.4$''$  &  15$'' \times 2''$ \\
&&& \\
\hline
\end{tabular}
\vskip1truecm
 
In Fig.~7, we present the four spectrographs (two slits, two channels).
Along Slit M and Slit N, significant signals were detected around 4 positions
(M1, M2, M3, M4) and 5 positions (Na, Nb, Nc, Nd, Ne), respectively.
They are listed in Table 4. The approximate positions of these sources
are also marked in Fig.~6 and Fig.~7.  For each source, we have defined 
an aperture size, given in the last column of Table~4.  This aperture was
used to extract a 1-d spectrum.  The resulting blue and red spectra 
of these sources are plotted in Fig.~8 for Slit M and Fig.~9 for Slit N.

\subsection{Emission Lines}
%
%\begin{figure}
%\plotone{table_line_1.ps}
%\end{figure}
%\begin{figure}
%\plotone{table_line_2.ps}
%\end{figure}

In Table 5 we list the following measured quantities for
the detected lines:
\begin{description}
\item{(1)} the corresponding redshift in km s$^{-1}$;
\item{(2)} velocity dispersion (in km s$^{-1}$);
\item{(3)} line flux in $10^{-15}$ erg s$^{-1}$ cm$^{-2}$.
\end{description}

\vskip1truecm
\noindent{\bf Table 5. Emission lines}
\nopagebreak

\hskip-0.5truecm\begin{tabular}{lcccccccccc}\hline
          &  \multicolumn{2}{c} {~~~~~~~~~~M1~~~~~~~~~~~}  &      M2  &        M3  &    M4 &     Na   &     Nb  &      Nc  &    Nd  &   Ne  \\
          &    comp.~1  & comp.~2        &  &            &       &	     &	       &	  &	   &	   \\   
\hline
$[$SII$]$ $\lambda$6731 & &	 &	     &		  &	  &	     &	       &	  &	   &       \\
V (km s$^{-1}$)  & 6686 &  &    6017 &            &       &    6197  &    5756 &     6102 &   6693 &  6246 \\
$\delta$V (km s$^{-1}$)   &  381 & &      368 &            &       &    1010  &    1115 &     1341 &    297 &   499 \\
flux$^{\dagger}$ & 0.94 & &      0.89 &            &       & 2.05$^{\ddagger}$  & 1.40$^{\ddagger}$ &     1.68$^{\ddagger}$ &   0.18 &  0.62 \\
          &      &    &       & 		  &	  &	     &	       &	  &	   &       \\ 
$[$SII$]$ $\lambda$6717 & &	 &	     &		  &	  &	     &	       &	  &	   &       \\ 
V (km s$^{-1}$)  & 6700 &  5987&    6030 &       &   6253&    6164  &    5713 &     6020 &   6697 &  6221 \\
$\delta$V (km s$^{-1}$)   &  366 & 346 &      375 &       &    405&     611  &     316 &      348 &    619 &   550 \\
flux$^{\dagger}$ & 1.29 & 0.16 &      1.20 &       &   0.11&    1.85$^{\ddagger}$  &    0.50$^{\ddagger}$ &     0.22$^{\ddagger}$ &   0.40 &  1.11 \\
          &      &         &  &            &       &  	     &         &	  &        &	   \\                   
$[$NII$]$ $\lambda$6583 & &	 &	     &		  &	  &	     &	       &	  &	   &	   \\
V (km s$^{-1}$)  & 6745 &    &  5970 &       6471 &       &    5889  &    5588 &     5927 &   6574 &  6251 \\
$\delta$V (km s$^{-1}$)   &  382 & &       429 &        513 &       &    1024  &    1110 &      703 &    573 &   515 \\
flux$^{\dagger}$ & 1.63 &    &  2.54 &       1.14 &       &    8.09$^{\ddagger}$  &    4.60$^{\ddagger}$ &     3.60$^{\ddagger}$ &   0.62 &  1.18 \\
          &      &          & &            &       &    	     &         &	  &        &	   \\                   
H$_\alpha$   &	 &	    & &		  &	  &	     &	       &	  &	   &	   \\
V (km s$^{-1}$)  & 6673 &    &  5988 &       6445 &   6125&    6152  &    5722 &     6060 &   6677 &  6249 \\
$\delta$V (km s$^{-1}$)   &  364 & &       357 &        433 &    342&     495  &     403 &      430 &    576 &   558 \\
flux$^{\dagger}$ & 10.0 &    &  5.64 &       1.27 &   0.24&    2.79$^{\ddagger}$  &    1.48$^{\ddagger}$ &     1.78$^{\ddagger}$ &   2.87 &  3.76 \\
          &      &           & &            &       & 	     &         & 	  &        &       \\             
$[$NII$]$ $\lambda$6548 & &	 &	     &		  &	  &	     &	       &	  &	   &       \\ 
V (km s$^{-1}$)  & 6689 &    &  6001 &       6414 &       &    6156  &    5695 &     6142 &   6682 &  6251 \\        
$\delta$V (km s$^{-1}$)   &  354 & &      309 &        426 &       &     596  &     807 &      288 &    345 &   364 \\        
flux$^{\dagger}$ & 0.91 &   &   0.58 &      0.37 &       &    0.80$^{\ddagger}$  &    0.56$^{\ddagger}$ &     0.16$^{\ddagger}$ &   0.58 &  0.41 \\
          &      &         &  &            &       &	     &	       &	  &	   &	   \\   
$[$OI$]$ $\lambda$6300  & &	 &	     &		  &	  &	     &	       &	  &	   &	   \\
V (km s$^{-1}$)  & 6761 &     &      &            &       &    6392  &    6317 &     6622 &   6676 &  6268 \\
$\delta$V (km s$^{-1}$)   &  587 & &           &            &       &    1134  &    1038 &     1268 &    557 &   623 \\
flux$^{\dagger}$ & 0.41 &    &       &            &       &    1.90  &    0.78 &     0.92 &   0.22 &  0.79 \\
%          &      &           &            &       &	     &	       &	  &	   &	   \\
%          &      &           &            &       &	     &	       &	  &	   &	   \\   
\hline
\end{tabular}

$^{\dagger}$ Flux of the line emission in 10$^{-15}$ erg s$^{-1}$cm$^{-2}$.

$^{\ddagger}$ Severely blended.
\vskip1truecm

\noindent{\bf Table 5. Continue}
\nopagebreak

\hskip-0.5truecm\begin{tabular}{lcccccccccc}\hline
          &  \multicolumn{2}{c} {~~~~~~~~~~M1~~~~~~~~~~~}  &      M2  &        M3  &    M4 &     Na   &     Nb  &      Nc  &    Nd  &   Ne  \\
          &    comp.~1  & comp.~2        &  &            &       &	     &	       &	  &	   &	   \\   
\hline
He~I $\lambda$5876  &	 &	&     &		  &	  &	     &	       &	  &	   &	   \\
V (km s$^{-1}$)  & 6688 &   &   6024 & 		  &	  &	     &	       &	  &	   &	   \\
$\delta$V (km s$^{-1}$)   &  383 & &       219 & 		  &	  &	     &	       &	  &	   &	   \\
flux$^{\dagger}$ & 0.35 &   &   0.19 & 		  &	  &	     &	       &	  &	   &	   \\
          &      &         &  & 		  &	  &	     &	       &	  &	   &	   \\
$[$OIII$]$ $\lambda$5007&	& &	     &		  &	  &	     &	       &	  &	   &	   \\
V (km s$^{-1}$)  & 6530 &5812 &   5812 &            &       &    6830  &    5512 &          &   6591 &       \\
$\delta$V (km s$^{-1}$)   &  303 & 269 &      264 &            &       &     298  &     206 &          &    276 &       \\
flux$^{\dagger}$ & 3.59 & 0.43  &   1.42 &            &       &    0.31  &    0.22 &          &   0.66 &       \\
          &      &          & & 		  &	  &	     &	       &	  &	   &	   \\
$[$OIII$]$ $\lambda$4959& &	 &	     &            &       &          &         &          &        &       \\ 
V (km s$^{-1}$)  & 6534 &    &  5868 &            &        &          &         &          &        &       \\         
$\delta$V (km s$^{-1}$)   &  287 & &      316 &            &       &          &         &          &        &       \\ 
flux$^{\dagger}$ & 1.05 &   &   0.38 &            &   &          &         &          &        &       \\         
          &      &         &  &            &       &     &         &          &        &       \\                 
H$_\beta$   &	 &   &  &		  &	  &	     &	       &	  &	   &	   \\
V (km s$^{-1}$)  & 6665 &  &    5986 &	     &	 &          &    5678 &          &   6789 &  6233 \\               
$\delta$V (km s$^{-1}$)   &  285 & &   342 & 	  &       &          &     316 &          &    422 &   339 \\      
flux$^{\dagger}$ & 1.45 &    &  1.49 & 		      &          &    & 0.37 &          &   0.44 &  0.83 \\          
          &      &        &   &            &             &         &          &        &       \\                  
H$_\gamma$   & 	 &	   &  &		  &	  &	     &	       &	  &	   &	   \\
V (km s$^{-1}$)  & 6635 &    &  6014 & 		  &	  &	     &	       &	  &	   &	   \\
$\delta$V (km s$^{-1}$)   &  279 & &      335 & 		  &	  &	     &	       &	  &	   &	   \\
flux$^{\dagger}$ & 0.39 &   &   0.67 & 		  &	  &	     &	       &	  &	   &	   \\
          &      &         &  &            &       &	     &	       &	  &	   &	   \\   
$[$OII$]$ $\lambda$3727 &	& &	     &		  &	  &	     &	       &	  &	   &       \\
V (km s$^{-1}$)  & 6600 &     5876&      &            &       &    6117  &    5796 &          &   6600 &  6198 \\
$\delta$V (km s$^{-1}$)   &  451 & 355&          &            &       &     526  &     283 &          &    937 &   487 \\
flux$^{\dagger}$ & 1.82 &    1.08&       &            &       &    2.31  &    0.82 &          &   1.48 &  2.34 \\
%          &      &           &            &       &	     &	       &	  &	   &	   \\   
\hline
\end{tabular}

$^{\dagger}$ Flux of the line emission in 10$^{-15}$ erg s$^{-1}$cm$^{-2}$.

\vskip1truecm

\setcounter{figure}{9}
\begin{figure}
%\vskip-1truecm
\vspace{4truecm}
%\plotone{fig10.ps}
%\vskip-0.8truecm
\caption{Close-up plots of the following lines:
[0II]$\lambda$3727, H$_\beta$, [OIII]$\lambda$4959/$\lambda$5007, [OI]$\lambda$6300,
H$_\alpha$/[NII]$\lambda$6548/$\lambda$6563, [SII]$\lambda$6716/$\lambda$6731.
}
\end{figure}
%\vskip1truecm

In Fig.~10 we present close-up plots of the following lines:
[0II]$\lambda$3727, H$_\beta$, [OIII]$\lambda$4959/$\lambda$5007, [OI]$\lambda$6300,
H$_\alpha$/[NII]$\lambda$6548/$\lambda$6563, [SII]$\lambda$6716/$\lambda$6731.

\subsection{Notes for Individual Sources} %\label{sect.4.3}

\begin{description}
\item{M1:} The brightest source, near the center of
the IGM starburst associated with the
MIR source SQ-A (Xu et al. 1999). The line emission is predominantly
from the 6600 km s$^{-1}$ component (comp.~1 in Table 5). 
The most conspicuous
evidence for the second component (comp.~2 in Table 5), centered at 
$\sim 5900$ km s$^{-1}$, is the second peak of the 
[OII]$\lambda$3727 line. This second component also shows
up in the [OIII]$\lambda$5007 line and the [SII]$\lambda$6716 line.
\item{M2:} The second brightest source, also associated with
the MIR source SQ-A. However, different from M1, the line
emission of this source is from the 6000 km s$^{-1}$ component.
The emission from the 6600 km s$^{-1}$ component is not detected here.
This is in good agreement with Moles et al. (1998).
\item{M3:} This is a relatively faint source, located
within the H$_\alpha$ bridge linking the shock front and 
the Seyfert-2 nucleus of NGC~7319. Only one component
centered at $\sim 6450$ km s$^{-1}$ is detected. The
marginal detection of the
[SII]$\lambda$6716 line at 6600 km s$^{-1}$ is deemed unreliable
because of the large errors associated with the skyline subtraction.
The high $I([NII]\; \lambda 6583) /I(H_\alpha)$
ratio (0.90) suggests that the H$_\alpha$ bridge is either
a cooling flow (Heckman et al. 1989), or
gas bow-shocked by an AGN jet (Aoki et al. (1999).
\item{M4:} Another weak source, associated with a diffuse
emission region in the periphery of the IGM starburst. The 
$I([SII] \lambda 7317)/I(H_\alpha)$ ratio indicates that
the emission is shock excited rather than
radiatively excited (Dopita \& Sutherland 1995). 
Apparently it is associated with the Source 5
in the Chandra image (Fig~1 of Trinchieri et al. 2003). 
\item{Na:} Located near the center of the shock front.
Three velocity systems (6800, 6100 and
6400 km s$^{-1}$) are detected in the emission lines,
with the 6100 km s$^{-1}$ component dominating the H$_\alpha$,
[NII], [SII] and [OII] lines, and the 6800 km s$^{-1}$ component
dominating the high excitation [OIII] $\lambda$5007 line.
The very broad $[OI]\; \lambda 6300$ line ($\delta V = 1134$ km s$^{-1}$),
which is the only line showing the 6400 km s$^{-1}$ component,
may well be due to the blending of the same line
in the 6100 and 6800 km s$^{-1}$ systems. The relatively
large velocity offset of the $[NII]\; \lambda 6583$ line
from 6100 km s$^{-1}$, and its large velocity dispersion
($\delta V = 1024$ km s$^{-1}$) may also be attributed to the contamination
of the H$_\alpha$ emission of the 6800 km s$^{-1}$ component.
From the line width of the [OIII] $\lambda$5007 line
and of the [OII] $\lambda$3727 line, the 6100 km s$^{-1}$
component has substantially higher velocity dispersion
($\delta V = 526$ km s$^{-1}$) than the 6800 km s$^{-1}$
component ($\delta V = 298$ km s$^{-1}$). 
\item{Nb:} Located at the intersection between the
shock front and a tidal tail of the intruder NGC~7318b.
Two velocity systems (6300 and
5700 km s$^{-1}$) are detected in the emission lines. 
The very broad $[OI]\; \lambda 6300$,
$[NII]\; \lambda 6583$ and $[SII]\; \lambda 6731$ lines
suggest possible blending with lines in another 
velocity system of V$\sim 6600$ km s$^{-1}$. The line width
of [OIII] $\lambda$5007, H$_\beta$
and [OII] $\lambda$3727 indicates that, without blending,
the velocity dispersion of the 5700 km s$^{-1}$ component
is $\sim 300$ km s$^{-1}$ (close to the velocity resolution).
\item{Nc:} Located immediately north of
Na, this is another source associated with the shock front.
Two velocity systems (6100 km s$^{-1}$ and 6600 km s$^{-1}$) are detected.
As in Na, the 6100 km s$^{-1}$ component dominates the
H$_\alpha$, [NII], and [SII] lines. But, unlike in Na, 
the central velocity of
the $[OI]\; \lambda 6300$ line (6622 km s$^{-1}$)
indicates that the high velocity component dominates 
this line emission, though the broad line width 
($\delta V = 1268$ km s$^{-1}$) suggests contribution
from the low velocity component.
\item{Nd:} Associated to the same starburst region as
M1, but off-set by $\sim 3''$ toward the east. Only one
component associated with the IGM (6600 km s$^{-1}$) is
detected. The [SII] lines are affected by sky-line
subtraction, therefore with relatively large uncertainties.
\item{Ne:} At the southern end of the shock front, likely
to be associated with the post-shock gas. Only one
component is detected, with the central velocity
of $\sim 6250$ km s$^{-1}$.
\end{description}

\section{Physical Conditions in the IGM of SQ}

\subsection{In the Ionized Gas}
\subsubsection{IGM Starburst (SQ-A) Region}
Four sources (M1, M2, M4 and Nd) in Table~5
are in this region. M1 and M2 are close to the
core of the IGM starburst, and M4 and Nd are in
the periphery. As indicated in previous HI,
CO and Fabry-P\`erot observations, there are two velocity
systems with recession velocities of
$\sim$6600 km s$^{-1}$ and $\sim$6000 km /sec, respectively, 
in this region.
The main component of M1 (comp.~1) and Nd belong 
to the 6600 km s$^{-1}$ system, while the second component of M1
and M2 belong to the 6000 km s$^{-1}$ system. M4, with a velocity
of $\sim 6200$ km s$^{-1}$, is perhaps associated with the
hot X-ray gas (see Section 4.3) and has no counterpart in the neutral gas.

We have detected most of the diagnostic lines
in both M1 and M2 with relatively high signal-to-noise
ratios. Both sources show typical HII-region line
ratios. In Table 6 we list the physical parameters
derived from the line ratios:

\vskip1truecm
\noindent{\bf Table 6. Physical Conditions in SQ-A}
\nopagebreak

\hskip-0.5truecm\begin{tabular}{lcccccccccc}\hline
 (1)   & (2)             & (3)   & (4)       & (5) & (6)    \\
source & $ 12+log[O/H]$  & $n_e$ & $T_{ion}$  & Q & A(H$_{\alpha}$) \\
       &                 & ( cm$^{-3}$ ) & ( K )       &   & ( mag )  \\
\hline
M1     &  8.76$\pm$0.06  & $< 100$ & $\sim 40000$ & $\sim 10^8$ & 2.0 \\
M2     &  8.95$\pm$0.09  & $< 100$ & $\sim 40000$ & $\sim 10^7$ & 0.64 \\
\hline
\end{tabular}

\begin{description}
\item{Notes:}
\item{column (2):} Metal abundance derived using the model of Kewley \&
Dopita (2002). To be compared with the Solar value 
$ 12+log[O/H] =8.69 \pm 0.05$ (Allende Prieto et al. 2001).
\item{column (3):} Density derived from $I([SII]\; \lambda 6717) /I([SII]\; \lambda 6731)$.
\item{column (4):} 
Ionization temperature derived, using the model of Evans \& Dopita
(1985), from two extinction-insensitive line ratios:
$I([OIII]\; \lambda 5007) /I(H_\beta\; \lambda 4861)$ and
$I([NII]\; \lambda 6584) /I(H_\alpha\; \lambda 6563)$.
\item{column (5):} 
Ionization parameter derived, using the model of Evans \& Dopita
(1985), from two extinction-insensitive line ratios:
$I([OIII]\; \lambda 5007) /I(H_\beta\; \lambda 4861)$ and
$I([NII]\; \lambda 6584) /I(H_\alpha\; \lambda 6563)$.
\item{column (6):} H$_\alpha$ extinction derived from
Balmer decrement, assuming the Case B intrinsic line ratio
$I(H_\alpha\; \lambda 6563)/I(H_\beta\; \lambda 4861) = 2.86$
and average Galactic interstellar extinction curve
(Savage and Mathis 1979).
\end{description}
\vskip1truecm

It appears that both cold gas systems, as revealed by
the data of M1 and M2, have the metal abundance
slightly above the solar value.
They both have density $< 100$ cm$^{-3}$ and $T_{ion}\sim 40000$ K.
These are in the range of typical HII regions. The ionization
parameter of M2 is an order of magnitude lower than that of
M1, indicating the former has a lower density
and/or larger distance from the ionizing stars. This is consistent
with the more extended morphology and much lower
extinction of M2 compared to M1. The extinction derived
from the Balmer decrement of M1 is A(H$_\alpha$)=2 mag, consistent with 
its rather red continuum color (Sulentic et al. 2001) and
the very compact morphology of the 6600 km s$^{-1}$ HI gas
(Williams et al. 2002).

\subsubsection{Shock-front Region}
Na, Nb, Nc and Ne are in the shock front region.
They all have 
$I([OI]\; \lambda 6300) /I(H_\alpha+[NII]) > 0.1$ and
$I([SII]\; \lambda 6717,6713) /I(H_\alpha+[NII]) > 0.25$,
therefore are dominantly shock excited (Dopita \& Sutherland 1995).
The kinematics of the ionized gas in these sources
is rather complex, but consistent with previous results
from the HI and Fabry-P\'erot observations (Sulentic et al. 2001).
In  Na and Nc, which are north of
NGC~7318b, most of the emission lines are dominantly
due to the low velocity (6000 km s$^{-1}$) system.
Indeed, the HI observations (Williams et al. 2002) detected
only the 6000 km s$^{-1}$ component in this region. 
The high velocity system (6600 km s$^{-1}$), with which
the low velocity HI gas is colliding, shows up only in
[OI] $\lambda$6300, and [OIII] $\lambda$5007 lines.

The fact that the 6600 km s$^{-1}$ component is undetected in HI 
in shock front suggests that most of it has
been processed by the shock and, consequently, been
converted to hot gas or ionized gas (Sulentic et al. 2001).
In the framework of the models of Dopita \& Sutherland (1995),
this means that the 6600 km s$^{-1}$ component does not
have the precursor HII region in front of the shock front.
This may be one of the reasons why 6600 km s$^{-1}$ component is 
relatively faint in these regions 
because, without the precursor, its
Balmer lines are about a factor of 2.3 fainter
(Dopita \& Sutherland 1995).

Because of the severe blending among [SII] and [NII] lines,
it is difficult to derive the density and other physical
parameters using line ratios. Nevertheless, given the
close relation between the ionized gas in the
shock front and in SQ-A, it is likely that the 6600 km s$^{-1}$
and 6000 km s$^{-1}$ components have the metal abundance of
M1 and M2, respectively. From the extended
morphology of the H$_\alpha$ emission (Xu et al. 1999)
and of the 6000 km s$^{-1}$ HI gas in
the shock front, one can safely infer a rather
low gas density (much lower than that in M1),
possibly in the range of $n_e \sim 0.01$ --- 0.1~cm$^{-3}$ 
as derived by Trinchieri et al. (2003) from the
X-ray and HI data. For a gas temperature of 5.8 $10^6$ K 
(Trinchieri et al. 2003) and a head-on collision ($\phi =90^\circ$)
between the clouds, the shock velocity is
$v_{shock}\sim 460 ~\,{\rm km~s}^{-1}$ 
(Eq.~1 of Trinchieri et al. 2003). This is
very close to the shock velocity derived from
the {\it line-of-sight} relative velocity 
($400~\,{\rm km~s}^{-1} = 4/3\times 600/2~\,{\rm km~s}^{-1}$),
confirming that the direction of the 
collision is nearly parallel to the line-of-sight 
(Sulentic et al. 2001).
 
The non-detection of the $H_\beta$ emission in Na corresponds to
an upperlimit of I(H$_\beta$) $< 0.3$ 10$^{-15}$ erg~s$^{-1}$~cm$^{-2}$.
Assuming an intrinsic I(H$_\alpha$)/I(H$_\beta$)=3 (Dopita \& Sutherland
1995), a very high extinction (E(H$_\beta$-H$_\alpha) >$ 1.2) is hinted 
at when comparing the $H_\beta$ upperlimit with the $H_\alpha$ flux
reported in Table 5. Indeed, substantial IR emission is
detected around Na region (Fig.~2). 

Nb and Ne are in the south of NGC~7318b. 
 Nb is very close (within a few arcsec)
to the two 'unusual emission line regions' (U1 and U2) observed by Ohyama
et al. (1998). Nb,
U1 and U2 all show lines consistent with shock excitation
models. However, the velocities of
U1 and U2 are 6560 km s$^{-1}$ and 6729 km s$^{-1}$ (Ohyama et al. 1998), 
while the emission lines (except for the [OI] line) 
in Nb are dominated by the 5700 km s$^{-1}$ component.
The spectroscopic observations of
Gallagher et al. (2001) for a region near Nb 
also confirm that the co-existence of both the 5700 km s$^{-1}$
and 6600 km s$^{-1}$ emission line systems.
It is likely that these
emission line regions associated with
different gas systems are ionized by the UV light 
generated by the same shock that is induced by their collision.

Located near the south end of the shock front,
Ne looks different from Na, Nb and Nc, in the sense that only
one velocity system is detected, and none
of the emission lines shows the very broad (1000+ km s$^{-1}$)
line width commonly found in the other three sources.
In addition, the velocity of Ne (V = 6250 km s$^{-1}$) is 
inconsistent with either 5700 km s$^{-1}$ or 6600km s$^{-1}$,
the two velocity systems found in the HI gas around
this position. It is very likely that this emission region
is due to post-shock gas which is now cooling through
emission lines. Its [SII] line ratio (1.79 with an uncertainty
of 20\%) and Balmer decrement are consistent with a low density gas 
of moderate extinction (E(H$_\beta$-H$_\alpha) = 0.45$).

\subsection{Magnetic Field in the Shock-front Region} 

The ridge radio source, which is associated with
the shock front, emits $L_{\rm 1.40~GHz} \approx 3 \times
10^{22}$~W~Hz$^{-1}$ from a volume extending about $90'' \times 15''$
($\sim 10^{23}$~cm $\times 2 \times 10^{22}$~cm) on the sky
plane.  To estimate the minimum-energy (equipartition) magnetic field
strength, we assume that its depth along the line of sight is also
$\sim 2 \times 10^{22}$~cm, the filling factor is unity, and the
relativistic ion/electron energy ratio is $k \approx 40$, the value
for cosmic rays in our Galaxy.  The minimum-energy field is $H_{\rm
min} \approx 10\,\mu$G.  The corresponding energy density of
relativistic particles plus magnetic fields is $U_{\rm min} \approx
1.0 \times 10^{-11}$~erg~cm$^{-3}$.  This is significantly lower than
the IGM thermal energy density implied by the X-ray emission, so the
radio source is dynamically insignificant.  Its observed morphology
is but a passive reflection of the X-ray and [N II] distributions.
The lack of detectable polarization in the ridge suggests that the
ridge magnetic fields may be disordered, 
as expected if their configuration is determined by external forces.
Both "beam depolarization"
and Faraday rotation may have caused the reduced polarization.

The ridge radio luminosity exceeds by at least a factor of ten the
radio luminosities of all star-forming regions in SQ (NGC 7318a, SQ-A,
SQ-B, plus any possible non-AGN emission from NGC 7319).  If it were
produced by an intracluster starburst as proposed by van der Hulst \&
Rots (1981), the expected star-formation rate would be about $30
M_\odot$~yr$^{-1}$ and the expected free-free emission at 1.4~GHz would be
about 10\% of the total (Condon 1992).  The optical continuum from
such a luminous starburst is not seen, and our spectroscopic data
and the $H_\alpha$ 
luminosity of the ridge rule out such a luminous thermal radio
component.  Rather, we believe that the high ridge luminosity supports
the argument that shocks accelerated the radiating electrons, although
some of the seed electrons might have been stripped from one or more
normal SQ galaxies in the past.

\subsection{Dust Emission in the Shock-front Region}
The morphological resemblance between the ridge of FIR emission in
the residual map (Fig.~3) and the diffuse ridge of radio
synchrotron emission (Fig.~4) is striking.
Thus, the FIR emission in the shock region
might also be associated with the passage
of the shock front.
If a sufficient fraction of the refractory elements
in the colliding clouds is in the solid state, the plasma immediately
downstream of
the shock will primarily cool by inelastic collisions of electrons and
ions
on dust grains. Dwek \& Werner (1981) showed that for the conditions
of the interstellar medium in the solar neighborhood, this cooling
mechanism dominates over cooling via inelastic collisions
between electron and ions if the gas temperature immediately
downstream of the shock exceeds about $10^{6}~$K.
For the X-ray ridge in SQ, Trinchieri et al. (2003)
found a temperature of 0.5 KeV (5.8 $10^{6}~$K).

Another necessary condition for the dust emission to dominate
the cooling of a shock is that dust grains can survive
the sputtering during the cooling.
For gas temperatures between $10^{6}$ and $10^{9}$~K, Draine \& Salpeter
(1979) give the sputtering
time scale for a spherical grain of radius $a$ embedded in a plasma
of hydrogen density $n_{\rm H}$ as:
\begin{eqnarray}
t_{\rm {sput}} \sim 10^{6}
\left [ \frac{a}{\mu{m}} \right ]
\left [ \frac{n_{\rm {H}}}{{\rm cm}^{-3}} \right ]^{-1}\,\,\,\,{\rm yr}
\end{eqnarray}
For $n_{\rm H}\,=\,0.027\,{\rm cm}^{-3}$ 
(hot gas, Trinchieri et al. 2003) and
$a\,=\,0.1\,\mu{m}$, $t_{\rm {sput}}$ is $3.7\times10^{6}~$yr. This can
be
compared with the gas cooling time scale $t_{\rm {cool}}$ due to
collisions with
grains. For $a\,=\,0.1\,\mu{m}$, a gas temperature of $5.8~10^{6}~$K,
$n_{\rm H}\,=\,0.027\,{\rm cm}^{-3}$, and a dust-to-gas mass ratio of 0.006,
$t_{\rm {cool}}$ is $\sim\,2.1\times10^{6}~$yr (Dwek \& Werner 1981). Thus,
if the dust-to-gas mass ratio in the intergalactic medium of SQ is
of the same order as that for the ISM in the Milky Way,
the sputtering time scale will be comparable to the cooling
time scale, so a significant fraction of the dust grains
will survive and cool the gas.
As noticed by Trinchieri et al. 2003), 
the distribution of the hot gas is likely to be clumpy
and the actual density can be significantly higher than the mean
($n_{\rm H}\,=\,0.027\,{\rm cm}^{-3}$). However,
since $t_{\rm {cool}}$ and $t_{\rm{sput}}$
have very similar dependencies on $n_{\rm H}$ and $a$,
the ratio $t_{\rm {sput}}/t_{\rm {cool}}$ should not vary much with
these quantities.

The cooling mechanism due to dust emission is rather efficient
(Dwek and Werner 1981). For
sufficiently fast shocks it can tap almost all the kinetic energy
flux flowing through the shock, which,
following Dopita \& Sutherland (1996), is:
\begin{eqnarray}
L_{\rm T} & = & 2.1\times10^{42}
\left [ \frac{v_{\rm shock}}{460~{\rm km~s}^{-1}} \right ]^{3.0}
\left [ \frac{n_{\rm{o}}}{0.01 {\rm cm}^{-3}} \right ]
\left [ \frac{A}{100{\rm kpc}^{-2}} \right ]\,\,\,{\rm
erg~s}^{-1}~{\rm cm}^{-2}
\end{eqnarray}
\noindent where $v_{\rm shock}$ ($\sim 460{\rm km~s}^{-1}$) 
is the shock velocity, $A$ is the area of the
shock and $n_o$ the upstream number density
(for a strong shock, $n_o\sim n_{\rm H}/4$). For the case of the colliding
HI clouds in SQ, $n_{\rm o}$ and A can be estimated to be
$\sim\,$0.01$\,{\rm cm}^{-3}$ and $\sim\,$100~kpc$^{2}$, respectively. 
These values are obtained from the typical HI column density of 
3\,10$^{20}\,{\rm cm}^{-3}$ (see Fig.~5
of Sulentic et al. 2001) and by assuming the line of sight depth of the
shock to
be equal to the 10~kpc extent of the radio ridge seen in the plane of
the sky.

The resulting value of 2.1$\times10^{42}$~erg~s$^{-1}$ 
for $L_{T}$ from Eq.~4 is to be compared with the observed FIR 
luminosity of the shock region. In Table~2 this region
is bound together with the binary NGC~7318b and NGC~7318a,
with a total $f_{60\mu m}=76$ mJy and $f_{100\mu m}=230$ mJy. 
The higher resolution 15$\mu m$ and 11.4$\mu m$ maps suggest
that roughly $\sim 30\%$ of the FIR emission is due to
the shock front, the rest is contributed by
the binaries (the two nuclei plus the giant star formation 
region in the south of NGC~7318b, see Fig.~2).
We calculate the FIR luminosity of the shock region by scaling the
predicted form of the SED (Fig.~11) to 
a value of 69~mJy at 100~${\mu m}$ (estimated as 30 percent of
the flux density of the combined 100$\mu m$ emission of 230 mJy
from the shock and N7318a/b),
and integrating over frequency. Assuming the distance of
80~Mpc, this yields a total dust luminosity
$L_{{\rm dust}}=1.9~10^{42} {\rm erg~s}^{-1}$, comparable 
to the $L_{{\rm T}}$ of 2.1$\times10^{42}$~erg~s$^{-1}$ derived from Eq.~4. 
It should also be noted that $L_{{\rm dust}}$ is about
an order of magnitude higher than that of
the X-ray luminosity of the ridge:
$L_{{\rm X}}\sim 1.5~10^{41}
{\rm erg~s}^{-1}~{\rm cm}^{-2}$ (Trinchieri e al.
2003), supporting the argument that the dust emission
is the dominant cooling mechanism for the shock.
%Fig.~1:
\begin{figure}
%\vskip-1truecm
\vspace{2truecm}
%\plotone{fig11.ps}
%\vskip-0.8truecm
\caption{The model SED of collisionally heated dust emission
in the SQ shock front region. It has 
been scaled to a flux density of 69 mJy at a wavelength of
100 micron.
In the model (Popescu et al. 2000), the spherical grains in the size 
range 0.001 to 0.25${\mu m}$ are stochastically heated in a plasma of solar
abundances with $n_{\rm H}\,=\,0.027\,{\rm cm}^{-3}$. The ion and 
electron temperatures were each taken to be 5.8 $10^{6}~$K. The grains 
were taken to be astronomical silicate and graphite with relative
abundances taken from Draine \& Lee (1984), and
with a distribution in number density over size $a$ 
proportional to $a^{-2.5}$. This size distribution is flatter
than the interstellar size distribution to take into account the
effects of sputtering, as discussed in Popescu et al. (2000). The 
predicted SED peaks at around 120~${\mu m}$.
}
\end{figure}

In the cooling time scale of $2.1\times 10^6$ yr, a shock of velocity
$460~{\rm km~s}^{-1}$ will move a distance of 1.0 kpc, or 2.5 arcsec
at the distance of SQ. Therefore, the dust emission should trace the
shock structure very closely, as predicted by Popescu et al. (2000)
for the case of accretion shocks in clusters of galaxies. In the MIR
maps (Fig.~2) where the shock front can be discerned from other
sources, it is indeed unresolved perpendicular to the axis of the
ridge. The predicted SED (Fig.~11) from shock heated dust has a flux
density ratio $f_{100 \mu m}/f_{60 \mu m}$ of 3.0, very close to the
observed color ratio of the combined dust emission 
of the shock and NGC~7318a/b
(Table 2), which is mostly due to photon heated dust.  This highlights
the point made by Popescu et al. (2000), in the context of FIR
emission from the intracluster medium when viewed from cosmological
large distances, that the intergalactic and galactic FIR emission
components will be difficult to distinguish on the basis of color,
despite the different dust heating mechanisms involved.

We conclude that the observed FIR emission in the shock front region
can be accounted for in terms of collisional heating of the grains by
the plasma immediately downstream of the shock, provided that the
refractory elements in the (pre-shock) IGM of SQ are mainly in the
solid state, and the dust abundance in the HI clouds is at least as
much as that in the interstellar medium of the Milky Way.

\section{SQ-A: Star Formation Rate and Triggering Mechanism}

\subsection{Star Formation Rate}

From the H$_\alpha$ luminosity (uncorrected for
the extinction) and the 15$\mu m$ luminosity
of SQ-A, Xu et al. (1999) found that the star formation rate is
0.66 M$_\sun$ yr$^{-1}$ (H$_\alpha$) or 0.81 M$_\sun$ yr$^{-1}$ (15$\mu m$).
With the new spectroscopic
and FIR data, we can now better constrain these estimates.

First, we exploit the new spectroscopic information to improve the
determination of the H$_\alpha$ fluxes using the two narrow band
(H$_\alpha$+[NII] emission) images, one primarily samples the 6600 km
s$^{-1}$ component and the other the 5700/6000 km s$^{-1}$ component
(Xu et al. 1999).  Assuming the H$_\alpha$-to-[NII] line ratios and
the extinction corrections of the 6600 km s$^{-1}$ component are the
same as those of M1, and those of the 6000 km s$^{-1}$ component are
the same as those of M2, we obtain the H$_\alpha$ fluxes (before and
after extinction correction) and the H$_\alpha$ luminosities of the
two components in SQ-A (Table 7).  The total uncorrected H$_\alpha$
flux, 1.30 10$^{-13}$ erg s$^{-1}$ cm$^{-2}$, is in very good
agreement with the result of Xu et al. (1999) of 1.27 10$^{-13}$ erg
s$^{-1}$ cm$^{-2}$.  The star formation rate (SFR) can be derived from
the H$_\alpha$ luminosities using the STARBURST code (Leitherer et
al. 1999) with the following parameters: continuous star formation,
Salpeter IMF, $M_{low}=1\; M_\sun$, $M_{up}=100\; M_\sun$, age of the
starburst: 10 Myr.  The resulted total SFR (1.45 M$_\sun$ yr$^{-1}$,
see Table 7) is about a factor of 2 higher than that of Xu et
al. (1999), primarily due to the extinction correction.

\vskip1truecm

\noindent{\bf Table 7. H$_\alpha$ Luminosities and Star-formation Rate in SQ-A}
\nopagebreak

\hskip-0.5truecm\begin{tabular}{lcccccccccc}\hline
velocity components & f$_{H_\alpha}$ & f$_{H_\alpha}$ & L$_{H_\alpha}$
& S.F.R. \\
 & (no ext. corr.) & (ext. corr.) & (ext. corr.) & \\
& 10$^{-14}$ erg s$^{-1}$ cm$^{-2}$ &  10$^{-14}$ erg s$^{-1}$ cm$^{-2}$ 
&  $10^{40}$ erg s$^{-1}$ & M$_\sun$ yr$^{-1}$ \\
\hline
6600 km s$^{-1}$ & 8.31 & 52.4 & 40.6 & 1.25 \\
6000 km s$^{-1}$ & 4.65 & 8.38 & 6.43 & 0.20 \\
total            & 13.0 & 60.8 & 47.0 & 1.45 \\
\hline
\end{tabular}

\vskip1truecm

The star formation rate can also be estimated from the FIR
luminosity. The 60$\mu m$ and 100$\mu m$ flux densities of SQ-A (Table
2) correspond to a FIR luminosity of $L_{{rm 40--120}\mu m}=2.52\times
10^{42}$ erg s$^{-1}$.  In order to obtain the total IR (5 ---
1000$\mu m$) luminosity $L_{IR}$, we estimate the total emission of
large grains using the median ratio of $L_{cd}/L_{{\rm 40--120}\mu m} =
2.06$ found by Popescu \& Tuffs (2002) for the cold dust emission in
Im-BCD galaxies, and approximate the emission of small grains/large
molecules by the 15$\mu m$ luminosity ($\nu L_\nu$) of SQ-A which is
$1.82\times 10^{42}$ erg s$^{-1}$ (Xu et al. 1999).
This results in $L_{IR}=7.01\times 10^{42}$ erg s$^{-1}$.
Then assuming 60 percent of the bolometric
luminosity of the starburst is absorbed by dust and re-emitted
in the IR (Xu et al. 1994), we obtain a total (bolometric) luminosity 
of the starburst of 1.17 10$^{43}$ erg s$^{-1}$.
Compared to the results of the same STARBURST simulation 
described above, the FIR luminosity
corresponds to a star formation rate of 0.33 M$_\sun$ yr$^{-1}$.
This is about a factor 4 lower than the star formation rate
derived from the H$_\alpha$ luminosities.

The discrepancy could be caused by the following factors:
\begin{description}
\item{(1)} Possible over-correction of the H$_\alpha$ extinction of the
6600 km s$^{-1}$ component which is estimated using the Balmer
decrement of M1. This could happen if M1, 
located at the core of the starburst, 
has encountered more extinction than HII regions elsewhere in SQ-A.
However, this seems not the case because
the Balmer decrement of source Nd, 
which is at the periphery of the 6600 km s$^{-1}$ H$_\alpha$
emission region, gives an $A_{H_\alpha}= 1.9$ mag, 
in good agreement with that of M1 (2.0 mag).  Furthermore,
even without any extinction correction, the SFR estimated
from the H$_\alpha$ luminosity is already a factor of
$\sim 2$ higher that that derived from the FIR luminosity.
Therefore, the discrepancy is unlikely to be caused
by over-correction of the H$_\alpha$ extinction of the
6600 km s$^{-1}$ component. 
\item{(2)} Errors in the $f_{60\mu m}$ and $f_{100\mu m}$.
A hint for this can be drawn from the relatively low
$f_{60\mu m}/f_{15\mu m}$ ratio (=5.5) of SQ-A, which is about
a factor of 4 lower than the mean (=21.5$\pm 9.4$) 
of the eight closely interacting and starbursting galaxy pairs 
observed by Xu et al. (2000). This also explains why the
star formation rate derived using $f_{60\mu m}$ and $f_{100\mu m}$
is significantly lower than that derived using $f_{15\mu m}$
(Xu et al. 1999). Both $f_{60\mu m}$ and $f_{100\mu m}$
are derived after the subtraction of brighter
sources NGC~7319 and NGC~7320, so could be seriously affected by 
errors in the subtraction. 
\item{(3)} A top-heavy IMF. If the starburst has more massive 
ionizing stars than predicted by a Salpeter IMF, one will have
more H$_\alpha$ emission for a given bolometric luminosity.
There has been evidence that starbursts may tend to have
top-heavy IMFs (Rieke et al. 1980; Bernloehr 1993). Given the
unusual nature of SQ-A (an IGM starburst triggered by a high speed
collision), it will be very interesting to find out in future
studies whether indeed it has an unusual IMF.
\end{description}
Future FIR observations with higher angular resolutions than
ISOPHOT (e.g. SIRTF-MIPS observations) will distinguish the possibilities
(2) and (3).

\subsection{Triggering Mechanism}
Xu et al. (1999) summarized the supporting
evidence available by then for
the argument that SQ-A is an ongoing starburst. These include
(1) strong MIR emission, (2) high 
H$_\alpha$ equivalent-line-width ($> 100${\AA}),
(3) weak NIR (K band) emission. Dividing the stellar mass estimated from
the K-band luminosity by the star-formation rate estimated from the MIR and 
H$_\alpha$ luminosities, Xu et al. (1999) found the age of the starburst 
to be $\lsim 10$---20 Myr. The ``$\lsim$'' sign reflects the possibility that
some of the K-band emission is due to an underlying older stellar
population (stars
stripped from the member galaxies in previous galaxy-galaxy
close-encounters). New HST observations of Gallagher et al. (2001) 
indeed found many very young ($\sim 5$ --- 6 Myr) star clusters in the region.
Also, large amount of molecular gas 
( 5 10$^8$ --- 10$^9$ M$_\sun$) has been discovered in this region 
(Gao \& Xu 2001; Smith \& Struck 2001; Lisenfeld et al.
2002), consistent with an ongoing starburst.

Xu et al. (1999) argued that this starburst is triggered directly by
the collision between the intragroup cold gas (the 6600 km s$^{-1}$ component
of the HI) and the cold gas associated with the intruder galaxy
NGC~7318b (the 6000 km s$^{-1}$ component of the HI), and is {\it not}
triggered by any tidal effects between NGC~7318b and other SQ members,
so should {\it not} be treated as a `tidal dwarf' as suggested by
some other authors (e.g. Mendes de Oliveira et al. 2001). 
The supporting evidence for this argument includes:
\begin{description}
\item{(1)} The H$_\alpha$ data show that the star formation in SQ-A
is occurring both in the IGM (6600 km s$^{-1}$ component) and
in the ISM of the intruder (the 6000 km s$^{-1}$ component). In particular,
the fact that the H$_\alpha$ emission is dominated by the
6600 km s$^{-1}$ component, conclusively rules out any interpretation for
the starburst that involves only processes within the intruder NGC~7318b.
\item{(2)} The age of the starburst ($\sim 10$ Myr) is consistent with
the dynamical time scale of the high speed collision. The probability
that both the starburst and the collision are happening simultaneously 
within such a short time scale would be very low if the former is not
causally related to the latter.
\item{(3)} It is even more implausible that, in such a short time scale,
two separated 'tidal dwarf' starbursts, one associated with the IGM
(the 6600 km s$^{-1}$ component) and the other with ISM of the intruder
(the 6000 km s$^{-1}$ component), are simultaneously happening 
(within such a small sky region) together with the collision. Therefore,
the conclusion is almost inevitable that the three events (the collision,
the two velocity components of the starburst) are directly related,
as depicted in the collision induced starburst scenario (as opposite
to the 'tidal dwarf' scenario).
\end{description}

Given the high collision velocity ($\sim 600$ km s$^{-1}$) 
and the apparent link
between SQ-A and the shock front in the H$_\alpha$ map (Fig.~6),
a candidate for the triggering mechanism of the starburst is
the star formation induced by a shock, as originally modeled
by Elmegreen \& Elmegreen (1978). Such a mechanism invokes the gravitational
instability in the post-shock gas for
the triggering of the star formation, and has been
applied to star formation in the jet-induced emission line regions in
the high-z radio galaxy 4C 41.17 by Bicknell et al. (2000).
However, in SQ-A, we did not detect
a significant component in the H$_\alpha$ emission associated with
the {\it post-shock} gas, which should have redshift $\sim 6300$ km s$^{-1}$
(assuming that HI clouds in the 6000 km s$^{-1}$ and the 6600 km s$^{-1}$
systems have similar mass distributions).
Instead, given that the two H$_\alpha$ components are closely associated with
the two HI velocity systems detected in this region (Williams et al. 2002),
the star formation is apparently happening in the {\it pre-shock} 
gas (see Section 7.1 for arguments on why 6000 km s$^{-1}$ gas is 
not post-shock gas).

A better model can be drawn from the theory developed by
Jog \& Solomon (1992, hear after JS), initially aimed to 
explain the origin of the intense
starbursts seen in colliding, gas-rich, field spiral galaxies.
In that theory, it is assumed that the collisions between the HI clouds,
which have much larger filling factors than the molecular clouds,
of two colliding gas systems lead to the formation of a hot ($\sim 1$ keV)
ionized, high pressure remnant gas. 
The over-pressure due to this hot gas
causes a radiative shock compression of the outer layers of pre-existing
giant-molecular-clouds (GMCs), which are embedded in
the HI clouds before the collision. 
The 'squeezing' radiative shock lasts only about 10$^4$ years 
(the crossing time of individual HI clouds), too short to destroy a GMC
but long enough to trigger instabilities in the thin outer layers.
These layers become gravitationally unstable 
and a burst of massive star formation is ignited 
in the initially barely stable GMCs. 

It is interesting to note that Xu et al. (1999) discounted this theory
for the following considerations: 
(1) it was not clear whether there is any molecular gas in
SQ-A region because it is very rare for molecular gas to be seen so far away
($\gsim 20$ kpc) from galaxy centers;  and (2) according to Pietsch et
al. (1997) SQ-A was in an X-ray hole in the ROSAT map, hence the
signal for the hot remnant gas produced by the ongoing collision
between HI clouds was missing.  Since the publication of Xu et
al. (1999), several new observations have shed new light on above
issues, and lent support for the JS model. First of all, Gao \& Xu
(2000) made the first detection of molecular gas in SQ-A with high
angular resolution ($\sim 8"$) interferometric CO observations using
BIMA. This was later confirmed by single dish observations of Smith \&
Stuck (2001) and Lisenfeld et al. (2002). The mass of the detected
molecular gas is $\sim 10^9\; M_\sun$, and both velocity components
(6600 and 6000 km s$^{-1}$) were detected. Secondly, the new Chandra
X-ray image of Trinchieri et al.  (2003), which is much more sensitive
than the ROSAT maps, shows that there is indeed an X-ray source at the
position of the core of SQ-A (corresponding to the peak in the
H$_\alpha$ map of the 6600 km s$^{-1}$ component and the peak in the
radio continuum maps).

Indeed it is expected in the scenario depicted by
the JS model that the starburst 
should have two components with the same velocities as the 
colliding cold gas systems since the motions of the
pre-existing GMCs (within which the starburst is taking place)
are little affected by the collision. This is because
(1) the GMCs do not collide with each other due to 
very low filling factors (Jog \& Solomon 1992); 
and (2) given GMCs' very high density and
compact configuration, and the rather short time scale 
($\sim 10^4$ yrs) for collisions between individual HI clouds
(within which the GMCs are embedded), little momentum will be
transferred from the GMCs to surrounding low density gas during
the collision.  The compression radiative shock
by the remnant gas is symmetric, hence will not affact the momentum
of the GMC's, either.

In this scenrio, the starburst starts immediately
after the collision taking place. Following JS (see their
Eq.~17), the star formation rate can be estimated as:
\begin{equation}
SFR = 1\; M_\sun\; yr^{-1}\; \left({M_{mol}\over 10^9 M_\sun}\right)
      \left({f\over 0.2}\right) \left({\alpha_m\over 0.1}\right) 
      \left({SFE\over 0.5}\right) \left({t\over 10^7\; yr}\right) 
\end{equation}
where $M_{mol}\sim 10^9 M_\sun$ 
is the total molecular gas in this region,
$f\sim 0.2$ is the fraction of this gas that is participating in
the collision. It is conceivable that a large fraction of 
the two gas systems may miss the collision if they both are in
arm-like configurations (when the collision is not perfect).
$\alpha$ is another fraction,
defined by the mean ratio of the mass of the unstable layer 
to the total mass of a GMC which is participating in the collision.
JS found that typically $\alpha_m =0.1$. Following again
JS, we assume that the typical star formation efficiency for
the unstable GMC layer is $SFE=0.5$. The time for
the two gaseous systems to pass each other is $t \sim 10^7$ yr.
With these plausible parameters, the star formation rate derived
in Eq.~7 agrees very well with what is observed for SQ-A
(Table 7).

In principle, the triggering mechanism proposed 
for the starburst in SQ-A should also work in other regions 
in SQ where the large-scale collision 
is taking place. The condition is the availability of pre-existing
 GMCs. Indeed, in the
south of NGC~7318b (around the position of
RA(J2000)=22$^h$35$^m$59$^s$, Dec(J2000)=33$^\circ$57$'$30$''$),
where marginally significant evidence for the CO emission
could be found in the BIMA image (Gao \& Xu 2000),
young star clusters ($10^7$ yr, Gallagher
et al. 2001) and the H$_\alpha$ emission (Xu et al. 1999) indicate
current star formation. This region was detected with bright
X-ray emission (Pietsch et al. 1998; Sulentic et al. 2001;
Trinchieri et al. 2003),
and the long slit spectroscopic observations of Gallagher
et al. (2001) show both the 5700 km s$^{-1}$ and 6600 km s$^{-1}$
systems. These are consistent with that the star formation
in this region may also be triggered by the same mechanism
as modeled by JS. Gao \& Xu (2000) did not detect any
CO emission in the shock front.
This may explain why there is no conspicuous star formation,
either in the form of young clusters (Gallagher et al
2001) or as bright point sources in H$_\alpha$ images (Xu e al. 1999),
in this region. It should be noted that Lisenfeld et al. (2002)
reported detection of the CO emission in the shock front region.
However, given the large beam ($22''$) of their single dish
observations, it is not very certain that the emission
is really from the shock front.

\section{Discussion}

\subsection{Is the 6000 km s$^{-1}$ Component the Post-shock Gas?}

In this paper, we have accepted the suggestion of Moles et al. (1997)
that both the 5700 km s$^{-1}$ and 6000 km s$^{-1}$ HI gas components
belong to NGC~7318b, the velocity difference due to the
rotation. Williams et al. (2002) questioned such a scenario based on
the apparent separation between the two HI components, in both spatial
and velocity distributions.  Sulentic et al. (2001) disputed this
argument and further supported the suggestion of Moles et al. (1997).
They pointed out the connection of the two HI components through the
ionized gas in the shock front.  Also, the Fabry-P\'erot observations
(Sulentic et al. 2001) demonstrate that the velocity of the emission
line regions along an arc, which includes the shock front and links
both the 5700 km s$^{-1}$ and 6000 km s$^{-1}$ HI components, changes
continuously from $\sim 5700$ km s$^{-1}$ to $\sim 6000$ km s$^{-1}$,
indicating a kinematic connection between the two velocity systems.

An alternative picture, as suggested by Lisenfeld et al. (2002), is
that the 6000 km s$^{-1}$ gas is indeed linked to 5700 km s$^{-1}$
component, but the velocity difference is due to the current
interaction with the 6600 km s$^{-1}$ component, instead of the
internal rotation of the ISM of NGC~7318b. Since the time scale is too
short for any gravitational effects, the only way to accelerate the
5700 km s$^{-1}$ gas to 6000 km s$^{-1}$ during the current
collision is through the shock. Hence,
in sich a scenario, the 6000 km s$^{-1}$ component should be the
post-shock gas. However, our spectroscopic observations do not support
this hypothesis. In both the SQ-A region and the shock front region,
where there was no detection of the 5700 km s$^{-1}$ HI gas, there is
no evidence for a 5700 km s$^{-1}$ component in the ionized gas,
either.  If all the 6000 km s$^{-1}$ cold gas (HI, molecular, and the
ionized) were processed gas already passed through the shock front,
then there would be nothing left for the 6600 km s$^{-1}$ gas to
collide with. Our conclusion is that the 6000 km s$^{-1}$ component is
not the post-shock gas.

\subsection{Distribution of Pre-shock Gas}

Nevertheless, the 5700 km s$^{-1}$ and 6000 km s$^{-1}$ components of
the HI gas do show peculiarities: (1) they are outside the main body of
NGC~7318b; (2) their maps do not show a rotating-disk morphology, even
after adding the ionized gas in the shock front; (3) in particular,
the 6000 km s$^{-1}$ component appears to be a round, extended cloud
centered at SQ-A (Fig.~9 of Williams et al. 2002), with little sign of
any substructure.

Moles et al. (1997) and Sulentic et al. (2001) argued that NGC~7318b
had been a 'normal' gas-rich galaxy before it entered SQ. In this
scenario, it is difficult to explain the above peculiarities of the HI
gas that is presumably associated with the galaxy. There are also
`abnormal' features in its optical morphology: (1) its outer optical
disk is one-side loped toward the north; (2) it has several long, open
arms which are usually seen in interacting galaxies (the so called
`tidal tails'). Williams et al. (2002) argued that
NGC~7318b is perhaps not entering the SQ the first time. However,
its relative velocity is too high for it to be gravitationally
bound to the SQ system, and therefore repeat passages seem to be
unlikely.  

Another possibility is that these
abnormal features are due to interaction with the elliptical galaxy
NGC~7318a, which has to be at least $\sim
100$ kpc behind NGC~7318b at the moment, although the projected
distance is only $\sim 10$ kpc. The requirement on the large
 line-of-sight distance between NGC~7318a and NGC~7318b
is because the time scale for
tidal effects is a few 100 Myr. Within this time, with the projected
relative velocity ($\delta V \simeq 900$ km s$^{-1}$), NGC~7318b would
have moved a few 100 kpc away from NGC~7318a since the close
encounter.  This also requires that NGC~7318b moves almost along the
line of the sight, otherwise the projected distance between NGC~7318a
and NGC~7318b would be much larger. Indeed, from the narrow width of
the shock front, and from the fact that in the shock front and in
SQ-A, emission line systems with velocities of $\sim 6600$ km s$^{-1}$
and of $\sim 6000$ km s$^{-1}$ are found on top of each other,
Sulentic et al. (2001) argued that the direction of the relative
velocity of NGC~7318b must be close to the line of sight.  Future
works such as higher sensitivity and higher resolution HI maps and
detailed theoretical simulations will help to solve the puzzle related
to the HI and optical morphology of NGC~7318b.

As shown by Sulentic et al. (2001), the pre-shock 6600 km s$^{-1}$ cold
gas is in a huge arc (longer than 100 kpc), with the SQ-A in the
northwest tip. This gaseous arc is likely to be a tidal feature
related to a previous close encounter between member galaxies, which
could have happened a few 100 Myr ago (Moles et al. 1997; Sulentic et
al. 2001).  The concentration of the HI gas (Williams et al. 2002) and
the molecular gas (Gao \& Xu 2000) in SQ-A, which is more than 20 kpc
away from any galaxy center, may indeed have the similar origin as
those HI-knots observed along the tidal tails (particularly at tips of
tidal tails) of interacting galaxies (Hibbard \& von Gorkom 1996; Duc
et al. 2000; Braine et al. 2001), although the triggering mechanism of
the IGM starburst is different from those of star-forming `tidal
dwarfs' as modeled by Barnes, J.E. \& Hernquist (1992) and Elmegreen
et al. (1993).

\section{Conclusions}

The compact galaxy group Stephan's Quintet (SQ) 
provides a unique target in the local
universe for studying the effects of high velocity collisions ($\sim
1000$ km s$^{-1}$) between two systems rich in cold gas.  The two
phenomenal events currently taking place in the IGM of SQ, namely the
large scale shock ($\sim 40$ kpc) and the IGM starburst SQ-A (SFR =
1.45 M$_\sun$ yr$^{-1}$), are very likely to be triggered by the same
ongoing collision between the intruder galaxy NGC~7318b ($v=5700$ km s$^{-1}$)
and the IGM ($v = 6600$ km s$^{-1}$). 
In this paper, we provided new constraints on the physical conditions
in the IGM involved, and investigated the physical mechanism linking
these events with the collision using new far infrared (FIR) images
at 60$\mu m$ and 100$\mu m$ (ISOPHOT C-100 camera), radio continuum
images at 1.4 GHz (VLA B-array) and 4.86 GHz (VLA C-array), and
long-slit optical spectrographs (Palomar $200"$ telescope).

We found that the shock front, which appears as a radio ridge and
dominates the radio continuum emission of SQ, has a steep nonthermal
spectral index ($\alpha = 0.93\pm 0.13$). Its FIR-to-radio flux ratio
is extremely low ($q < 0.59$) compared to that of galaxies ($<q> =
2.3\pm 0.2$), consistent with the hypotheses that the relativistic
electrons responsible for the radio emission are accelerated by the
large scale shock (in contrast to the relativistic electrons in galaxy 
disks which are likely to be accelerated by SNRs).  Its
observed IR emission can be explained in terms of collisional heating
of dust grains by the plasma immediately downstream of the shock.  The
long-slit spectra of sources in this region have typical emission line
ratios of shock-excited gas. The very broad line line widths 
(FWHM $\geq 1000$ km
s$^{-1}$), and the fact that in some cases more than two velocity
systems were detected along the same line of sight, provide further
evidence for an ongoing collision in this region.  The magnetic field
strength estimated using the minimum-energy assumption is 
$\approx 10\,\mu$G. The implied energy density of the electro-magnetic
field is significantly lower than the IGM thermal energy density
derived from the X-ray emission, indicating a minor role played by the
electro-magnetic force on the dynamics in the shock front.  No
linearly polarized emission brighter than 50~$\mu$Jy~beam$^{-1}$ was
found in any component of SQ at either 1.40 or 4.86~GHz, indicating
that the magnetic fields may be disordered, and both "beam depolarization"
and Faraday rotation may have caused the reduced polarization.

The IGM starburst SQ-A was clearly detected in both the FIR and the
radio bands. Its radio spectral index ($\alpha =0.8\pm 0.3$) and the
FIR-to-radio ratio ($q=2.0\pm 0.4$) are consistent with those of star
formation regions.  The optical spectra of two sources in this region,
M1 ($v=6600$ km s$^{-1}$) and M2 ($v=6000$ km s$^{-1}$), have typical
line ratios of HII regions. The metallicity of M1 is
$ 12+log[O/H] =8.76 \pm 0.06$ and that of M2 is 
$ 12+log[O/H] =8.95 \pm 0.09$, both being slightly higher  
than the solar value ($12+log[O/H] =8.69 \pm 0.05$,
Allende Prieto et al. 2001). This result
confirms that the IGM is stripped gas from galaxies and
rules out the possibility that it is primordial.
According to the Balmer decrement of M1,
the $6600$ km s$^{-1}$ component of the IGM starburst is heavily
obscured ($A_\alpha = 2.0$), while the Balmer decrement of M2
indicates a moderate extinction ($A_\alpha = 0.64$) for the $6000$
km s$^{-1}$ component.  The star formation rate estimated from
the extinction-corrected H$_\alpha$ luminosity of SQ-A is 1.45 M$_\sun$
yr$^{-1}$, of which 1.25 M$_\sun$ yr$^{-1}$ due to the $6600$ km
s$^{-1}$ component and 0.20 M$_\sun$ yr$^{-1}$ due to the $6000$ km
s$^{-1}$ component. The SFR estimated using the FIR luminosity is
significantly lower (SFR=0.33 M$_\sun$). 
The discrepancy is due either to errors in the
FIR flux densities or to a top-heavy IMF.  The very good agreement
in velocity between H$_\alpha$ and HI
(Williams et al. 2002) for both components 
 suggests strongly that the starburst
is occurring in the pre-shock gas rather than in the post-shock gas.
A model (Jog \& Solomon 1992) based on a scenario in which a starburst
is triggered in the outer layers of pre-existing (pre-shock) GMCs
compressed by surrounding shocked gas (which is the HI gas before the
collision), can be applied to SQ-A and explain the observed SFR, the two
components of the starburst, and the apparent link between SQ-A and
the large scale shock front.

%%%%%%%%%%%%%%%%%%%%%%%%%%%%%%%%%%%%%%%%%%%%%%%%%%%%%%%%%%%%%%%%%%%%%%%
% Acknowledgment:
%%%%%%%%%%%%%%%%%%%%%%%%%%%%%%%%%%%%%%%%%%%%%%%%%%%%%%%%%%%%%%%%%%%%%%%

\vskip1cm 
This research has made use of the NASA/IPAC Extragalactic Database
(NED) which is operated by the Jet Propulsion Laboratory, California
Institute of Technology, under contract with the National Aeronautics
and Space Administration. We thank Cristina Popescu and Jack Sulentic 
for helpful discussions.
%\vskip1truecm

%%%%%%%%%%%%%%%%%%%%%%%%%%%%%%%%%%%%%%%%%%%%%%%%%%%%%%%%%%%%%%%%%
% The references section.
%%%%%%%%%%%%%%%%%%%%%%%%%%%%%%%%%%%%%%%%%%%%%%%%%%%%%%%%%%%%%%%%%

\end{document}